\documentclass[aos,preprint]{imsart}

\RequirePackage[OT1]{fontenc}
\RequirePackage{amsthm,amsmath}
\RequirePackage{natbib}
\usepackage{color}
\usepackage{natbib} 
\usepackage{bm}
\usepackage{booktabs} 
\usepackage{amsmath,amssymb}
\usepackage{enumerate}
\usepackage{rotating}
\usepackage{color}
\usepackage{soul}

\numberwithin{equation}{section}
\theoremstyle{plain}

\newcommand{\ttt}{{\boldsymbol\theta}}

\newcommand*{\Cdot}{\raisebox{-0.25ex}{\scalebox{1.5}{$\cdot$}}}

\newcommand{\bfzero}{{\mathbf{0}}}
\newcommand{\bfone}{\mbox{$\mathbf 1$}}

\newcommand{\aaa}{{\boldsymbol\alpha}}

\newcommand{\aalpha}{{\boldsymbol\alpha}}

\newcommand{\ee}{{\mathbf e}}

\def\gg{\mbox{$\mathbf g$}}

\newcommand{\mm}{\mbox{$\mathbf m$}}

\newcommand{\pp}{\mbox{$\mathbf p$}}

\newcommand{\qq}{{\mathbf q}}
\newcommand{\rr}{{\mathbf r}}

\newcommand{\RR}{{\mathbf R}}

\def\sb{\mbox{$\mathbf s$}}

\def\uu{\mbox{$\mathbf u$}}

\def\vv{\mbox{$\mathbf v$}}

\newcommand{\xxi}{{\boldsymbol\xi}}

\newtheorem{theorem}{Theorem}
\newtheorem{definition}{Definition}
\newtheorem{example}{Example}
\newtheorem{lemma}{Lemma}
\newtheorem{proposition}{Proposition}
\newtheorem{remark}{Remark}

\begin{document}

\begin{frontmatter}
\title{Identifiability of restricted latent class models with binary responses}
\runtitle{Identifiability of restricted latent class models }

\begin{aug}
\author{\fnms{Gongjun} \snm{Xu}\ead[label=e1]{xuxxx360@umn.edu}}

\runauthor{G. Xu}

\affiliation{ University of Minnesota}

\address{Gongjun Xu\\
School of Statistics,\\
University of Minnesota\\
224 Church Street SE\\
Minneapolis, MN, 55455\\
\printead{e1}\\
%
}
\end{aug}

\begin{abstract}
Statistical latent class models are widely used in social and psychological researches,  yet it is often difficult to establish the identifiability of the model parameters.
In this paper we consider the identifiability issue of a family of restricted latent class models,
where the restriction structures are needed to reflect pre-specified  assumptions on the related assessment. 
We establish the identifiability results in the strict sense and specify which types of restriction structure would give the identifiability of the model parameters. The results not only guarantee the validity of many of the popularly used models, but also  provide a guideline for the related experimental design, where in the current applications the design is usually experience based and identifiability is not guaranteed. 
Theoretically, we develop a new technique to establish the identifiability result, which may be extended to other restricted latent class models.
\end{abstract}


\end{frontmatter}

\section{Introduction}

Statistical latent class models are widely used in social and psychological researches to model latent traits that are not directly measurable, with the aim to identify homogeneous subgroups of individuals based on their surrogate response variables. 
Although latent class models have many attractive traits for practitioners, 
fundamental identifiability issues,
 i.e., the feasibility of recovering model parameters based on the observed data, could be difficult to address.   
Specifically, we say a set of parameters $\beta$ for a family of distributions $\{f(x|\beta): \beta\in B\}$ is identifiable if distinct values of $\beta$ correspond to distinct probability density functions, i.e., for any $\beta$ there is no $\tilde\beta\in B\backslash\{\beta\}$ for which $f(x|\beta)\equiv f(x|\tilde\beta).$
Identifiability is the prerequisite for most common statistical inferences, especially parameter estimation, and its study dates back to \cite{Koopmans} and \cite{Koopmans50};
 see also \cite{McHugh,rothenberg1971identification,GOODMAN74,gabrielsen1978consistency} for further developments.

For latent class models with finite mixtures of finite measure products, 
\cite{teicher1967identifiability} established the equivalence between the model identifiability with that of the corresponding one dimensional mixture model. 
 \cite*{gyllenberg1994non} further showed that the latent class models with binary responses (finite mixture of  Bernoulli products) are not identifiable. 
 Such nonidentifiablity results have likely impeded statisticians from looking further into this problem \citep*{allman2009identifiability}. 
 Recently, researchers have considered the generic identifiability of such models. 
The generic identifiability is defined following algebraic geometry terminology. It implies that the set of parameters for which the identifiability does not hold has Lebesgue measure zero.   
Establishing the identifiability conditions can be mathematically difficult. 
The generic identifiability problem is closely related to the algebraic geometry theory, as pointed out  by  \cite*{elmore2005application}. 
 { \cite{elmore2005application} and \cite{allman2009identifiability} used algebraic-geometric approaches to establish generic identifiability results for a large set of models, including the latent class models and many other latent variable models.}
In particular, the work of \cite{allman2009identifiability} is based on the fundamental result of Kruskal's trilinear decomposition  of three-way arrays \citep{kruskal1976more,kruskal1977three} by `unfolding' a high-way  array into a three-way array.

{The existing techniques to establish generic identifiability, being algebraic-geometric  in nature, necessarily exclude a measure zero  set. 
Therefore, they do not provide information as to whether the model parameters are identifiable for submodels with additional constraints, where the constrained parameter spaces usually falls in a measure zero set.
To develop the identifiability conditions for such restricted models, we need techniques to incorporate the additional constraints.}

In this paper, we consider a class of restricted latent class models with binary responses (finite mixture of Bernoulli products). 
The class of models has recently gained great interests in psychological and educational measurement, psychiatry and other research areas, 
where a classification-based decision needs to be made about an individual's latent traits,  based on his or her observed surrogate responses (to test problems, questionnaires, etc.). 
The model parameters are restricted via a pre-specified matrix (see Section 2.1 for more details) to reflect the diagnostic assumptions about the latent traits. 
In particular, when there is no restriction, the model becomes the unrestricted latent class model.
Differently from the unrestricted models, 
the restriction matrix provides important information for applications, and therefore the strict identifiability needs to be satisfied to guarantee the validity of the models under different parameter constraints. 
Although researchers have long been aware of the identifiability problem of  these types of restricted models \citep*{DiBello,MarisBechger,TatsuokaC09,deCarlo2011},  there is a tendency to gloss over the issue in practice due to a lack of theoretical development on the topic. To the author's best knowledge, there are few studies in the literature on the identifiability of the restricted latent class models.

This paper aims to address the identifiability issue for these models.
Our main contribution includes the following points. 
\begin{itemize} 
\item[i)] First, we prove the identifiability for a class of restricted latent class models.
We show the identifiability depends on the structure matrix and propose a unified set of sufficient conditions under which the model parameters are estimable from the data.
{For the restricted latent class models under consideration, the identification results are strict.}
From an application perspective, the identifiability results would provide a guideline for designing diagnostic tests, where in the current applications the design is usually experience based and the identifiability is often not guaranteed.
\item[ii)] Second, we develop a new technique to establish the identifiability results for a class of restricted latent class models. 
Instead of working on the tensor product, we propose to study the corresponding marginal matrix, which has a nice algebra structure that can be well incorporated with the specified constraints.
\end{itemize}

The remainder of this paper is organized as follows. Section 2 introduces the class of restricted models and contains useful background on  the diagnostic classification modeling and applications. 
Section 3 introduces the issue of identifiability and our main results.
The corresponding proofs are given in Section 4.

\section{Models and Applications}

\subsection{Model setup}\label{sec:Model}
The models begin from the basic setting, in which subjects (examinees, patients, etc) provide  a $J$-dimensional binary response vector $\RR = (R_1,...,R_J)^\top$ to $J$ items (test questions, symptom diagnostic questions, etc), {where the superscript $\top$ denotes the transpose,} and these responses depend in certain way on $K$ unobserved latent traits (attributes, skills,  etc). 
A complete set of $K$ latent traits is known as a latent class or an attribute profile,
which is denoted by column vectors $\aalpha= (\alpha_1,\ldots, \alpha_K)^\top$, where $\alpha_k \in \{0,1\}$ indicate the absence or presence, respectively, of the $k$th attribute.
The above structure of $\aaa$ is often assumed in psychological and educational measurement for the diagnosis purpose. 
For instance, in a diagnostic math exam, teachers aim to estimate whether a student has mastered certain math skills; in a psychiatry diagnosis, doctors want to know whether a patient has certain mental depressions.  
 Both $\aalpha$ and $\RR$ are subject-specific; a particular subject $i$'s attribute and response vectors are denoted by $\aalpha_i$ and $\RR_i$, respectively, for $i = 1,\ldots, N$. 
 We  assume that the subjects are a random sample of size $N$ from a designated population so that their attribute profiles $\aalpha_i$, $i=1,..., N$ are i.i.d.\ random variables following a multinomial distribution with probabilities
\begin{equation*}
P(\aalpha_i = \aalpha)= p_{\aalpha},
\end{equation*}
where $p_{\aalpha}\in (0,1)$, for any $\aalpha\in\{0,1\}^{K}$, and $\sum_{\aalpha}p_{\aalpha}=1$.
The distribution is thus characterized by the column vector $\pp=(p_{\aalpha}: \aalpha\in \{0,1\}^K)^\top$.

Given a subject's attribute profile $\aaa$, the response $R_j$ to item $j$ under the corresponding model follows a Bernoulli distribution
\begin{equation} \label{prob}
P(R_j = r\mid \aaa) = (\theta_{j,\aaa})^{r}(1-\theta_{j,\aaa})^{1-r}, \quad r=0, 1,
\end{equation}
where we denote
 $$\theta_{j,\aaa}= P(R_j=1 \mid \aaa),$$ 
 which is the probability of providing positive response to item $j$ for subjects with $\aaa$.
Let $\Theta = (\theta_{j,\aaa})$ be a $J\times 2^K$ matrix containing the $\theta$ parameters. 
The unknown model parameters of the latent class model include $\Theta$ and $\pp$.

In the following, we write $\ee_i$ as a standard basis vector, whose $i$th element is one and the rest are zero.  We write $\mathbf 0$ and $\mathbf 1$ as the zero and one column vectors, i.e., $(0,...,0)^\top$ and $(1,...,1)^\top$, respectively. When there is no ambiguity, we omit the index of length.

We consider a class of restricted latent class models where
parameters $\Theta = (\theta_{j,\aaa})$ are constrained by 
 the relationship between the $J$ items and the $K$ latent traits.
 Such relationship is specified through a $Q$-matrix, 
 which is defined as a $J\times K$ binary matrix 
with entries $q_{jk}\in\{0,1\}$ indicating the absence or presence, respectively, of a link between the $j$th item and the $k$th latent trait.
The row vectors, $\qq_j$ of $Q$ correspond to the full attribute requirements of each item.
Given an attribute profile $\aalpha$ and a $Q$-matrix $Q$, we 
write $$\aaa\succeq \qq_j ~\mbox{ if }~ \alpha_k \geq q_{jk} \mbox{ for any }  k \in\{1, \ldots, K\},$$
and 
$$\aaa\nsucceq \qq_j ~\mbox{ if there exists $k$ such that } \alpha_k < q_{jk};$$
similarly we define the operations $\preceq$ and $\npreceq$.

 If $\aaa\succeq \qq_j$, a subject with $\aaa$ has all the attributes for item $j$ specified by the $Q$-matrix and would be most ``capable'' to provide a positive answer; on the other hand, if $\aaa'\nsucceq \qq_j$, the subject with $\aaa'$ misses some related attribute and is expected not to have a higher positive response probability than  
$\aaa\succeq \qq_j$.
In addition,  subjects without mastery of any latent traits ($\aaa=\bfzero$) is expected to have the lowest positive response probability.
Such constraints on $\Theta$ are proposed through the following monotonicity relations:
\begin{eqnarray}\label{assumption1}
 \max\limits_{\aaa:\,\aaa\succeq \qq_j} \theta_{j,\aaa}
=\min\limits_{\aaa: \,\aaa\succeq \qq_j} \theta_{j,\aaa} 
\geq \theta_{j,\aaa'}\geq \theta_{j,\bfzero},~\mbox{ for any }\aaa';
\end{eqnarray}
in addition, for any  $k\in\{1,\cdots,K\}$ and item $j$ with $\qq_j=\ee_k$, 
\begin{eqnarray}\label{assumption1'}
 \theta_{j,\mathbf 1}
> \max\limits_{\aaa:\, \aaa\nsucceq\ee_k} \theta_{j,\aaa}.
\end{eqnarray}

Assumption \eqref{assumption1} requires that, all the most capable subjects with $\aaa\succeq\qq_j$ have the same positive response probability. 
Assumption \eqref{assumption1'} assumes that for an item only requiring the $k$th attribute, the most capable subjects with $\aaa=\mathbf 1$ have higher positive response probability than those not having the $k$th attribute. 
Both assumptions are satisfied by many of the restricted latent class models as introduced in Section 2.2. 

The $Q$-matrix is the key part of the restricted diagnostic models and its structure makes them distinguished  from the unrestricted latent class models in the  literature. 
Since some $\theta$'s are restricted to be equal, the parameter space then falls in a measure zero set with respect to the whole parameter space under the unrestricted model.

\subsection{Examples and Applications}
The restricted latent class models in Section 2.1 have recently gained great interests in cognitive diagnosis with applications in educational assessment, psychiatric evaluation, and many other disciplines \citep*{Rupp,Tatsuoka}, where they are often called as  diagnostic classification models or  cognitive diagnostic models. 
Cognitive diagnosis  is the process of arriving at a classification-based decision about an individual's latent traits,  based on his or her observed surrogate responses. 
Measuring students' growth and success means obtaining diagnostic information about their skill set; this is very important for constructing efficient, focused remedial strategies for improving student and teacher results. 
The introduced models are important statistical tools developed in cognitive diagnosis 
to detect the presence or absence of multiple fine-grained skills or attributes. 

We use a simple example for an illustration of the model setup.
\begin{example}
Suppose that we are interested in testing two latent traits: addition and multiplication. 
Consider a test containing three problems and admitting the following $Q$-matrix,
\begin{equation}\label{exampleQ}
Q=\quad
\begin{tabular}{ccc}
\hline & addition & multiplication\\ \hline
$2+1$ & $1$ & $0$ \\
$3\times 2$ & $0$ & $1$ \\
$(2+1)\times 2$ & $1$ & $1$\\\hline
\end{tabular}
\end{equation}
We have four latent classes $\aaa =(0,0), (1,0), (0,1), $ and $(1,1)$, corresponding to subjects who do not master either addition or multiplication, who master only addition, who master only multiplication, and who master both, respectively. 
Take the first item for an example. 
Under the restrictions in  \eqref{assumption1} and \eqref{assumption1'}, subjects who master addition, $\aaa=(1,0)$, have a higher correct response probability than those who do not master addition, $\aaa=(0,0)$ or $(0,1)$; on the other hand, they have the same correct response probability as those who master both, $\aaa=(1,1)$, since the first item only needs addition.
\end{example}

The restriction structure in Section 2.1 is satisfied by many of diagnostic models.
An incomplete list of the popularly used restricted latent class models developed in recent decades includes the DINA (Deterministic Input, Noisy `And' gate) and NIDA (Noisy Inputs, Deterministic `And' gate) models \citep{Junker,dela}, the reparameterized unified/fusion model (RUM) \citep*{DiBello,Hartz},  
the DINO (Deterministic Input, Noisy `Or' gate) and NIDO (Noisy Inputs, Deterministic `Or' gate) \citep*{Templin}, the rule space method \citep*{Tatsuoka1983,Tatsuoka}, the attribute hierarchy method \citep*{AHM},  the Generalized DINA models \citep{dela2011}, 
and the general diagnostic model \citep*{davier2008general}; see also  \cite{HensonTemplin09} and \cite{Rupp}.
We use the following examples to introduce some of the popularly used models.

\begin{example}[DINA model]\label{DINA}
 The DINA model \citep{Junker} assumes a conjunctive relationship among attributes. That is, it is necessary to possess all the attributes indicated by the $Q$-matrix to be capable of providing a positive response. In addition, having additional unnecessary attributes does not compensate for the lack of necessary attributes.
For  item $j$ and attribute vector $\aaa$, we define the ideal response
$\xi_{j,\aaa}^{DINA}= I (\aaa \succeq \qq_j)$. 
The uncertainty is further incorporated at the item level, using the slipping and guessing parameters $\sb$ and $\gg$.
For each item $j$, the slipping parameter $s_j = P(R_j = 0\mid \xi_{j,\aaa}^{DINA} = 1)$ denotes the probability of the respondent making a negative  response despite mastering all necessary skills; similarly, the guessing parameter $g_j = P(R_j = 1\mid  \xi_{j,\aaa}^{DINA} = 0)$ denotes the probability of a positive response despite an incorrect ideal response. 
The response probability $\theta_{j,\aaa}$ then takes the form
\begin{equation}\label{DDINA}
\theta_{j,\aaa} = (1-s_j)^{\xi_{j,\aaa}^{DINA}}g_j^{1-\xi_{j,\aaa}^{DINA}}.
\end{equation}
 In this case, assumptions \eqref{assumption1} and \eqref{assumption1'} are equivalent to $1-s_j>g_j$ for any item $j$, which is usually assumed in applications.
\end{example}

\begin{example}[DINO model]\label{DINO}

In contrast to the  DINA model, the  DINO model assumes a non-conjunctive relationship among attributes, that is, one only needs to have one of the required attributes to be capable of providing a positive response.
The ideal response of the DINO model is given  by
$\xi_{j,\aaa}^{DINO} =   I(\alpha_k \geq q_{jk}\mbox{ for at least one }k).$
Similar to the DINA model, there are two parameters $s$ and $g$ for each item, and
$$\theta_{j,\aaa} = (1-s_j)^{\xi_{j,\aaa}^{DINO}}g_j^{1-\xi_{j,\aaa}^{DINO}}.$$
Again, assumptions \eqref{assumption1} and \eqref{assumption1'} are satisfied if $1-s_j>g_j$ for any $j$.
\end{example}

\begin{example}[G-DINA model]\label{GDINA}
\cite{dela2011} generalizes the DINA model to the
G-DINA model.
The  formulation of the G-DINA model based on $\theta_{j,\aaa}$ can be decomposed into the sum of the effects due the presence of specific attributes and their interactions. Specifically,
\begin{eqnarray*}
\theta_{j,\aaa}  &=&
\beta_{j0} 
+ \sum_{k=1}^{K}\beta_{jk}(q_{jk}\alpha_{k})
+ \sum_{k'=k+1}^{K}\sum_{k=1}^{K-1}\beta_{jkk'}(q_{jk}\alpha_{k})(q_{jk'}\alpha_{k'})\\
&&+\cdots + \beta_{j12\cdots K}\prod_{k}(q_{jk}\alpha_{k}).
\end{eqnarray*}
Note that  not all $\beta$'s in the above equation are included in the model. For instance, when $\qq_j \neq \mathbf 1^\top$, we do not need parameter $ \beta_{j12\cdots K}$ since $\prod_{k}(q_{jk}\alpha_{k})=0$.
To interpret, $\beta_{j0}$ represents probability of a positive response when none of the required attributes is present;
 when $q_{jk}=1$, $\beta_{jk}$ is included in the model and it shows the change in the positive response probability as a result of mastering a single attribute $\alpha_k$; 
when $q_{jk}=q_{jk'}=1$, $\beta_{jkk'}$ is in the model and it shows the change in the positive response probability due to the interaction effect of mastery of both 
$\alpha_k$ and $\alpha_{k'}$; 
similarly, when $\qq_j=\mathbf 1^\top$, $\beta_{j12\cdots K}$ represents the change in the positive response probability due to the interaction effect of  mastery of all the required attributes. 
Note that the assumption in \eqref{assumption1}, ${\max\limits}_{\aaa:\,\aaa\succeq \qq_j} \theta_{j,\aaa}
={\min\limits}_{\aaa: \,\aaa\succeq \qq_j} \theta_{j,\aaa}$, is automatically satisfied from the model definition from. 
\end{example}

\begin{example}[Linear logistic model and logit-CDM]\label{CRUM}
The linear logistic model \cite[LLM, see][]{jacques1993loglinear,maris1999estimating}
 is given by
\begin{equation}\label{crum}
\theta_{j,\aaa} = \frac{\exp(\beta_{j0}+\sum_{k=1}^K\beta_{jk}q_{jk}\alpha_{k})}{1+\exp(\beta_{j0}
+ \sum_{k=1}^K\beta_{jk}q_{jk}\alpha_{k})}.
\end{equation}
Equivalently 
$$\normalfont{\mbox{logit}}~\theta_{j,\aaa} = \beta_{j0}+\sum_{k=1}^K\beta_{jk}q_{jk}\alpha_{k}.$$
This is also called the compensatory reparameterized unified model (C-RUM).
The LLM model  (\ref{crum})  is recognized as a structure in multidimensional item response theory model or in factor analysis.
Again, we have ${\max\limits}_{\aaa:\,\aaa\succeq \qq_j} \theta_{j,\aaa}
={\min\limits}_{\aaa: \,\aaa\succeq \qq_j} \theta_{j,\aaa}$ from (\ref{crum}). 
\end{example}

\begin{example}[Reduced RUM model and log-CDM]\label{RUM}
Under the reduced version of the Reparameterized Unified Model \citep[Reduced RUM, see][]{DiBello,Rupp}, we have \begin{equation}\label{aaaa}
\theta_{j,\aaa} = \pi_j \prod_{k=1}^K{r_{j,k}}^{q_{jk}(1-\alpha_k)},
\end{equation}
where $\pi_j$ is the positive response probability for subjects who possess all required attributes and $r_{j,k}$, $0<r_{j,k}<1$, is the penalty parameter for not possessing the $k$th attribute. 
 Note that 
 the model is equivalent to the log-link model
 $$\log\theta_{j,\aaa}  =\beta_{j0} + \sum_{k=1}^{K}\beta_{jk}(q_{jk}\alpha_{k}).$$
 For the reduced RUM in \eqref{aaaa}, it is easy to see that assumptions \eqref{assumption1} and \eqref{assumption1'} are satisfied  by the definition.
\end{example}

Psychometricians have long been aware of the identifiability issue of the $Q$-matrix based latent class models \citep{DiBello, TatsuokaC09, deCarlo2011, MarisBechger}. 
For these models, identifiability affects the classification of respondents according to their latent traits, which is dependent on the accuracy of the parameter estimates.  
Unprincipled use of standard diagnostic models may lead to misleading conclusions about the respondents' latent traits \citep{MarisBechger, TatsuokaC09}. 
In the literature, the identifiability issue of diagnostic models has only been studied for some specific models.  
Recently  \cite{xu2013statistical}, \citet{Chen2014} and \cite{Xu15} studied the identifiability of the slipping and guessing parameters under the DINA model in Example \ref{DINA}. However, their technique highly depends on the assumption that the subjects with $\xi^{DINA}=0$ having the same response probability (i.e., the guessing parameters) and therefore cannot be applied to the general diagnostic models considered in this paper, where the $Q$-matrix restricted latent structure is more complicated.

\section{Main results}

We introduce the identifiability results in this section. 
Throughout the rest of the discussion,
we let $M_{j,\Cdot}$ denote the $j$th row of a matrix $M$ and $M_{\Cdot,k}$
   the $k$th column. 
 We write $\mathcal{I}_d$ as the $d\times d$ identity matrix.

\subsection{Identifiability and response marginal $T$-matrix}

The model parameters contain the parameter matrix $\Theta=(\theta_{j,\aaa})_{J\times 2^K}$ and proportion parameter $\pp=(p_{\aaa})_{2^K\times 1}.$
Note the joint distribution of $\RR$, conditional on the latent class $\aaa$,  is given by a 
$J$-dimensional $2\times\cdots\times 2$ table
$${\mathbb P}_{\aaa}(Q,\Theta) = \bigotimes_{j=1}^J
\begin{bmatrix} 
1-\theta_{j,\aaa}\\
\theta_{j,\aaa}
\end{bmatrix},$$
where the $\rr=(r_1,\cdots, r_J)$-entry of the table is 
\begin{equation}\label{Apr15_1}
\pi_{\rr,\aaa}(Q,\Theta)
=\prod_{j=1}^J (1-\theta_{j,\aaa})^{1-r_j}\theta_{j,\aaa}^{r_j}.
\end{equation}
Note that $\pi_{\rr,\aaa}(Q,\Theta)$ is the probability of observing $\rr$ given $Q,\Theta,$ and $\aaa$. 
Following the above notation, we can write  
$$ P(\RR=\rr\mid Q,\Theta,\pp) = \sum_{\aaa\in\{0,1\}^K} \pi_{\rr,\aaa}(Q,\Theta)p_{\aaa}. $$
We introduce the following identifiability definition for the $Q$-restricted latent class models in Section 2.1. 
\begin{definition}
We say that  $(\Theta,\pp)$ is identifiable if the following holds:
\begin{equation}\label{deff}
\forall \mathbf r,
 P(\mathbf R=\mathbf r \mid Q,\Theta,\pp) 
     =P(\mathbf R=\mathbf r \mid Q,\bar\Theta,\bar\pp)
    ~ \Longleftrightarrow ~
    (\Theta,\pp)=(\bar\Theta,\bar\pp) .
     \end{equation}
     \end{definition} 
Note that the above definition does not involve label swapping of the latent classes due to the fact that the labels of attributes are pre-specified from the knowledge of the $Q$-matrix.
On the other hand, for unrestricted latent class models,
the latent classes can be freely relabeled without changing the distribution of the data and the model parameters are therefore identifiable only up to label swapping.

To establish \eqref{deff} for the restricted latent models, directly working with the vectors  $P(\mathbf R=\mathbf r \mid Q,\Theta,\pp) $ is technically challenging. 
To better incorporate the induced  restrictions by the $Q$-matrix, we consider the marginal matrix as introduced in the following.

\paragraph{Marginal $T$-matrix}
The $T$-matrix $T(Q,\Theta)$ is defined as a $2^J\times 2^K$ matrix,
where the entries are indexed by row index $\rr\in \{0,1\}^J$ and column index $\aalpha$.
The $\rr=(r_1,\cdots, r_J)$th row and $\aaa$th column element of $T(Q,\Theta)$, denoted by $t_{\rr,\aalpha}(Q,\Theta)$,  is  
the marginal probability that a subject with attribute profile $\aalpha$ answers all items in subset
$\{j: r_j=1\}$ positively.
Thus $t_{\rr,\aalpha}(Q,\Theta)$ is the marginal probability that, given $Q,\Theta,\aalpha$, the random response $\RR\succeq\rr$, i.e.,
$$t_{\rr,\aalpha}(Q,\Theta) = P(\RR\succeq\rr\mid Q,\Theta,\aalpha).$$
When $\rr=\mathbf 0$, 
$t_{\bfzero,\aalpha}(Q,\Theta) = P(\RR\succeq \bfzero) = 1 \mbox{ for any } \aaa;$
and for any $\rr \neq \bfzero$, 
\begin{eqnarray*}
t_{\rr,\aalpha}(Q,\Theta) 
= \prod_{j: r_j = 1} P(R_j=r_j \mid Q,\Theta,\aalpha) 
= \sum_{\rr'\succeq \rr} \pi_{\rr',\aaa}(Q,\Theta).
\end{eqnarray*} 
In particular, for $\rr=\ee_j$ with $1\leq j\leq J$, $$t_{\ee_j,\aalpha}(Q,\Theta) =P(R_j=1 \mid Q,\Theta,\aalpha)=\theta_{j,\aaa}.$$
Let $T_{\rr,\Cdot}(Q,\Theta)$ be the row vector corresponding to $\rr$.
Then we know that for $j=1,\cdots, J$, 
$T_{\ee_j,\Cdot}(Q,\Theta) = \Theta_{j,\Cdot}.$
In addition, for any $\rr \neq \bfzero$, we can write
\begin{equation}\label{defT}
T_{\rr,\Cdot}(Q,\Theta) = \bigodot_{j: r_j=1} T_{\ee_j,\Cdot}(Q,\Theta),
\end{equation}
where $\odot$ is the element-wise product of the row vectors.

By definition, multiplying the $T$-matrix by the the distribution of attribute profiles $\pp$ results in a vector containing the marginal probabilities of successfully answering each subset of items correctly. The $\rr$th entry of this vector is
\begin{eqnarray*}
T_{\rr,\Cdot}(Q,\Theta)\pp = \sum_\aalpha t_{\rr,\aalpha}(Q,\Theta) p_\aalpha 
= P(\RR\succeq \rr\mid Q,\Theta,\pp).
\end{eqnarray*}

We can see that there is a one-to-one mapping between the  $T$-matrix and the vectors 
$P(\RR= \rr\mid Q,\Theta,\pp)$, $\rr\in\{0,1\}^J$. 
Therefore, \eqref{deff} directly implies the following proposition. 
\begin{proposition}\label{mle}
 $(\Theta,\pp)$ is identifiable if and only if 
for any  $(\bar\Theta,\bar\pp)\neq (\Theta,\pp)$, there exists $\rr \in \{ 0, 1\}^J$ 
such that 
\begin{equation}\label{gxx}
T_{\rr,\Cdot}(Q,\Theta) \pp\neq T_{\rr,\Cdot}(Q, \bar\Theta)\bar\pp.
\end{equation}
\end{proposition}

From Proposition 1, to show the identifiability of  $(\Theta,\pp)$, we only need to focus on the $T$-matrix and prove that if 
\begin{equation}\label{mleeq}
T(Q, \Theta)\pp=T(Q,\bar \Theta)\bar\pp,
\end{equation} 
then 
$\Theta =\bar\Theta$ and $\pp=\bar\pp$.
We will use this argument in the proof of the identifiability results.

\subsection{Identifiability results}
In this subsection, we present the main identifiability results. 
To illustrate which types of $Q$-matrix structure is required to satisfy \eqref{gxx}, we take as an example the basic DINA model  introduced in Example \ref{DINA}. We consider the ideal case where the $j$th response 
$R_j=\xi_{j,\aaa}$, where $\xi_{j,\aaa}$ denotes $\xi_{j,\aaa}^{DINA}$ as defined in the example. 
In this ideal case, $\theta_{j,\aaa}$ is known as $\xi_{j,\aaa}$ 
 and the only unknown parameter is $\pp$. 
Note that here $t_{\ee_j,\aaa}(Q,\Theta) = \xi_{j,\aaa}$ and
the identifiability condition is equivalent to 
\begin{equation}\label{apr10_1}
(\xxi_{j,\aalpha}; j=1,\cdots,J) \neq 
(\xxi_{j,\aalpha'}; j=1,\cdots,J)
\end{equation} for all $\aalpha\neq \aalpha'$.
Otherwise, if there exists $\aaa\neq\aaa'$ such that $(\xxi_{j,\aalpha}; j=1,\cdots,J) = 
(\xxi_{j,\aalpha'}; j=1,\cdots,J)$,  the corresponding columns of the $T$-matrix satisfy  
$T_{\Cdot,\aaa}(Q,\Theta)=T_{\Cdot,\aaa'}(Q,\Theta).$
This implies  the nonidentifiability of $\pp$. 

To guarantee \eqref{apr10_1},
the mathematical requirements on the $Q$-matrix structure for the ideal case
 are specified in the following definition. 
\begin{definition}\label{DefComp}
A $Q$-matrix is said to be \emph{complete} if $\{\ee_j^\top: j=1,...,K\}\subset \{\qq_j: j=1,\cdots,J\}$; otherwise, we say that $Q$ is \emph{incomplete}.
\end{definition}
\noindent To interpret, for each attribute there must exist an item requiring that and only that attribute. The $Q$-matrix is complete if there exist $K$ rows of $Q$ that can be ordered to form the $K$-dimensional identity matrix $\mathcal{I}_K$. A simple (and minimal) example of a complete $Q$-matrix is  the $K\times K$ identity matrix ${\cal I}_K$.
Completeness ensures that there is enough information in the response data for each attribute profile to have its own distinct ideal response vector. When a $Q$-matrix is incomplete, we can easily construct a non-identifiable example. 
For instance, consider the incomplete $Q$-matrix 
\begin{equation*}
Q=
\left(\begin{array}{cc}
1&1 \\
0&1
\end{array}\right).
\end{equation*}
The population parameter $\pp$ is non-identifiable in this case. Subjects with attribute profiles $\aalpha^1 = (1,0)^\top$ and $\aalpha^2 = (0,0)^\top$ have the same ideal responses, so \eqref{apr10_1} is not satisfied. It is easy to see that such argument holds for general incomplete $Q$-matrix. 

It has been established in the literature that the completeness of the $Q$-matrix 
is a sufficient and necessary condition for the identifiability of $\pp$
 in the ideal response case under DINA model with known $\Theta$ \citep{Chiu,Xu15}. 
For the diagnostic models with unknown $(\Theta,\pp)$,  completeness of the $Q$-matrix is not enough to guarantee the  identifiability  of $(\Theta,\pp)$.
For instance, \cite{Xu15} showed that, under the DINA model,
 a necessary condition for the identifiability of the guessing parameters, slipping parameters,  and $\pp$ is: (i) the $Q$-matrix is complete and (ii) each latent trait is required by at least three items. 

For diagnostic models in Section 2, we provide in the following a unified sufficient condition that ensures their identifiability.
Since the DINA model is a special case of the restricted latent class models, 
it is necessary that we need 
 to use a complete $Q$-matrix for the diagnostic models and we need at least three items for each attribute.
To establish identifiability for the general class of models, 
we list below the  conditions that will be used. 
\begin{enumerate}
\renewcommand{\theenumi}{C\arabic{enumi}}
\renewcommand{\labelenumi}{(\theenumi)}
\item \label{C1}
We assume that the $Q$-matrix takes the following form (after row swapping):
\begin{equation}\label{Qform}
Q=
\left(\begin{array}{c}
 {\cal I}_{K}  \\
  {\cal I}_{K}  \\
 Q'
\end{array}\right).
\end{equation} 
\item \label{C2} 
Suppose $Q$ has the structure defined in \eqref{Qform}. 
We assume that for any $k\in\{1,\cdots,K\}$, 
$(\theta_{j,\ee_k}; j> 2K)^\top \neq (\theta_{j,\mathbf 0}; j> 2K)^\top$.
That is, there exist at least one item in $Q'$ such that subjects with $\aaa=\ee_k$ have different positively response probability from that of subjects with $\aaa=\mathbf 0$. 
\end{enumerate}

\begin{remark}
Condition \ref{C1} is a little stronger than the necessity of the complete matrix    by requiring two such identify matrices. 
\ref{C1} itself implies that each attribute is required by at least two items.
We need such condition to ensure enough information to identify the model parameters for each attribute. 
Condition \ref{C2} is satisfied if we assume for $j>2K$,
$ \theta_{j,\mathbf 0}
< \min_{\aaa\neq \mathbf 0} \theta_{j,\aaa}.$ That is, for 
subjects without any latent traits, the positive response probability is the lowest among all latent classes. In practice condition C2  may be checked  by a   posteriori empirically after data have been collected. On the other hand, condition C2 is satisfied if $Q'$ can be written as (after row swapping):
$$
Q'=\begin{pmatrix}
\mathcal{I}_K\\
\cdots
\end{pmatrix}.
$$
Therefore, if there are three identity matrices in the $Q$-matrix, both C1 and C2 are satisfied. 
\end{remark}

Before stating the main theorem, we show in the following result that condition C1 itself  is not enough to establish the identifiability of  $(\Theta,\pp)$. 
\begin{proposition}\label{theorem0}
Under the model setup in Section 2.1,
there exist $Q$-matrices satisfying \ref{C1} but
 $(\Theta,\pp)$ is non-identifiable.
\end{proposition}
{The proof of Proposition \ref{theorem0} is given in the appendix.}
Our main identifiability result is as follows.
\begin{theorem}\label{theorem1}
Under the model setup in Section 2.1,
if conditions \ref{C1} and~\ref{C2} hold, $(\Theta,\pp)$ is  identifiable.
\end{theorem}

The theorem specifies the sufficient condition under which the restricted latent class model parameters $(\Theta,\pp)$ are identifiable from the response data.  
From an application perspective, the identifiability result would provide a guideline for designing diagnostic tests, where currently the design is usually experience based and may suffer identifiability problems.
In particular, for the diagnostic classification models introduced in Section 2, 
the model parameters are identifiable if the $Q$-matrix satisfies the proposed conditions \ref{C1} and~\ref{C2}. 
Therefore,  if single attribute items are possible, it is recommended to have at least two complete matrices in the test which guarantees \ref{C1}; moreover, from Remark 1, both \ref{C1} and~\ref{C2} hold if we have three identity matrices in the $Q$-matrix. 
The theoretical result would also help to improve existing diagnostic tests. 
For instance, when researchers find that the estimation results are problematic and the $Q$-matrix does not satisfy the identifiability conditions, it is then recommended to design new items such that the identifiability conditions \ref{C1} and~\ref{C2} are satisfied.

 When the identifiability conditions are satisfied, the maximum likelihood estimators of $\Theta$ and $\pp$ are consistent as the sample size $N\to\infty$. Specifically,
  we introduce  a $2^J$-dimensional response vector 
 {\small $\boldsymbol\gamma=\{
1,
{N}^{-1}\sum_{i=1}^NI(\RR_i \succeq \ee_1),
\cdots,
{N}^{-1}\sum_{i=1}^NI(\RR_i \succeq \ee_J),
{N}^{-1}\sum_{i=1}^NI(\RR_i \succeq \ee_1+\ee_2),
\cdots,
{N}^{-1}\sum_{i=1}^NI(\RR_i \succeq \mathbf 1)\}$}.
 From the definition of the $T$-matrix and the law of large numbers, we know
$\boldsymbol\gamma\to
T(Q,\Theta) \pp  
$ almost surely as $N\to\infty$. On the other hand, the maximum likelihood estimators $\hat\Theta$ and $\hat\pp$ satisfy 
\begin{equation*}
\|\boldsymbol\gamma - 
T(Q,\hat\Theta) \hat\pp\| \to 0 , 
\end{equation*}  where $\|\cdot\|$ is the $L_2$ norm.
Therefore,  
\begin{equation*}
\|T(Q,\Theta) \pp - 
T(Q,\hat\Theta) \hat\pp\| \to 0  
\end{equation*}  almost surely.  Then from the proof of Theorem 1, we can obtain the consistency result  that $(\hat\Theta,\hat\pp)\to (\Theta,\pp)$ almost surely. 
Furthermore, following a standard argument of the asymptotic  theory,
 we take Taylor's expansion of the loglikelihood function at  $(\Theta,\pp)$ and the central limit theorem gives the asymptotic normality of the estimators $(\hat\Theta,\hat\pp)$.

\begin{remark}   
It is worthwhile to mention that our proof is not based on the trilinear decomposition result in \cite{kruskal1976more}.
Kruskal's result is applied in \cite{allman2009identifiability} to show the generic identifiability up to label swapping. 
From their Corollary 5, a sufficient condition for the generic identifiability is that the number of items $J$ is at least $2K+1$.
Such a condition is weaker than \ref{C1} and \ref{C2} due to the fact that $C2$ implicitly requires a non-empty $Q'$ and thus $C1$ and $C2$ imply $J\geq 2K+1$. However, their result can not be directly applied for the $Q$-restricted latent class models.
In addition, we would like to point out that conditions \ref{C1} and \ref{C2} are different from the rank conditions required by Kruskal's result and may be weaker in some cases.
\end{remark}

\begin{remark}
When the $Q$-matrix is incomplete, the model parameters $(\Theta,\pp)$ are nonidentifiable. A particular case is when each row of the $Q$-matrix is $\mathbf 1^\top$, then the model becomes similar as the unrestricted latent class models with $2^K$ classes. In this case, 
generic identifiability results as in \cite{allman2009identifiability} can still be applied. For a general incomplete Q-matrix, such results are still unknown in the literature.
We plan to study the generic identifiability for the parameters in the constrained parameter space when the $Q$-matrix is incomplete. 
These results would be helpful for practitioners, especially when it becomes difficult or even impossible to design items with particular attribute specifications. 

        It is also possible in practice that there  exist certain hierarchical structures among the latent attributes. For instance, a certain attribute may be a prerequisite for other attributes. In this case, some $\pp$'s are restricted to be $0$. 
The method developed in this paper may be extended to this type of restricted latent class models, and we would like to study this in the future. 
\end{remark}

\section{Proof of the main results}

\subsection{Proof of Theorem \ref{theorem1}}

To show the identifiability, Proposition 1 implies that it suffices to show that for two sets of parameters $(\Theta,\pp)$ and $(\bar\Theta,\bar\pp)$ satisfying equation \eqref{mleeq}, 
we must have $(\Theta,\pp)=(\bar\Theta,\bar\pp)$. 

Without loss of generality, we arrange the  rows of $Q$ such that it takes the form of \eqref{Qform} in condition \ref{C1}.
For notational convenience, we write $t_{\ee_j,\aaa}(Q,\Theta)$ and  
$t_{\ee_j,\aaa}(Q,\bar\Theta)$
 as $t_{\ee_j,\aaa}$ and $\bar t_{\ee_j,\aaa}$, respectively. 
Note that by the definition of the $T$-matrix,
$t_{\ee_j,\aaa} =\theta_{j,\aaa}$ and $\bar t_{\ee_j,\aaa} = \bar\theta_{j,\aaa}$
for any $j\in\{1,\cdots,J\}$ and  $\aaa\in\{0,1\}^K$.
Therefore to show $\Theta=\bar\Theta$, it is equivalent to show 
$t_{\ee_j,\aaa}=\bar t_{\ee_j,\aaa}$ for any $j\in\{1,\cdots,J\}$ and $\aaa\in\{0,1\}^K$.

We prove the theorem in five Steps. Given equation \eqref{mleeq} that $T(Q,\Theta)\pp=T(Q,\bar\Theta)\bar\pp$,  we aim to prove the following conclusions in each step:
\begin{figure}[h!]
  \centering
    \includegraphics[width=0.8\textwidth]{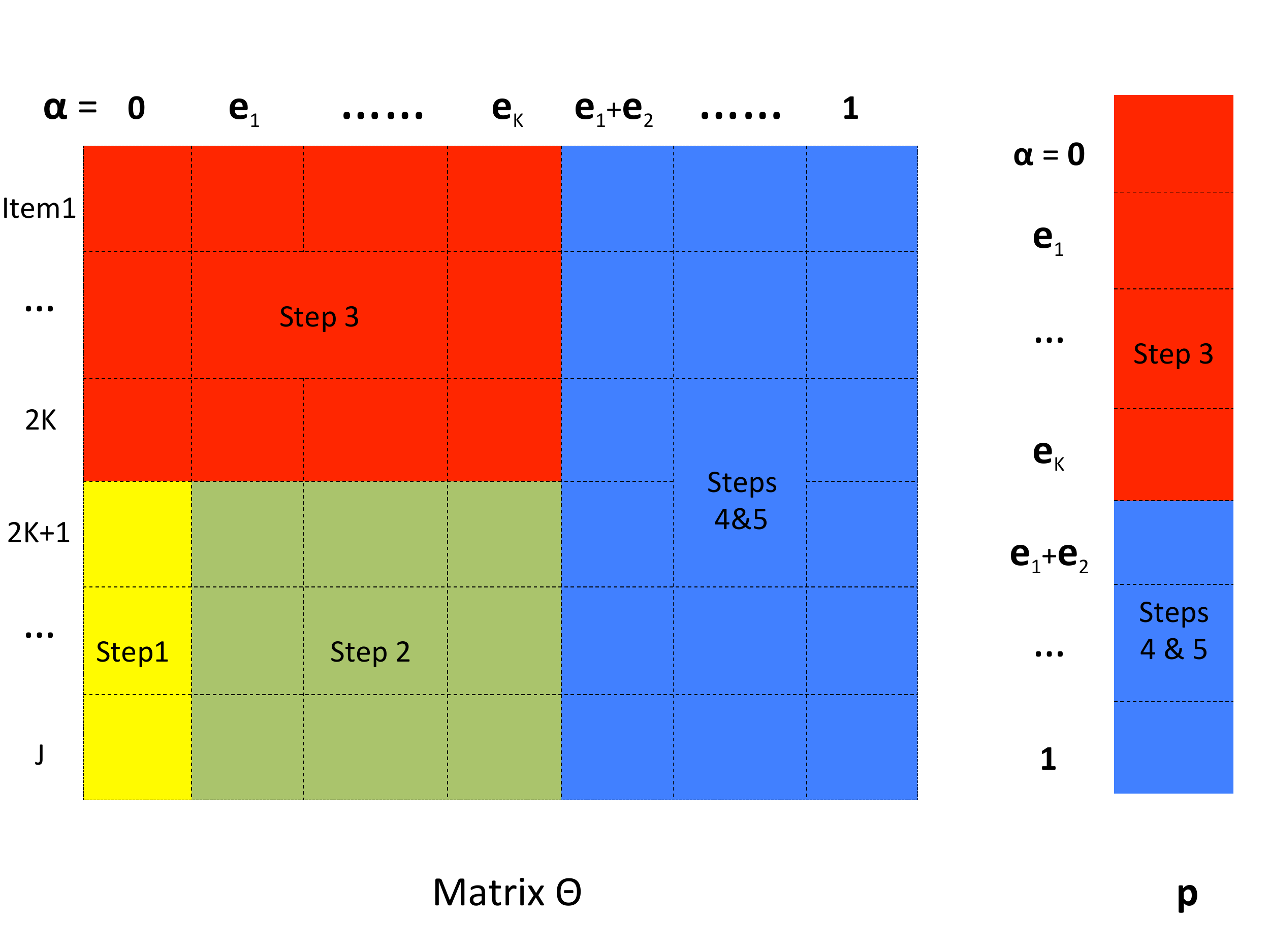}
      \caption{An illustration of the proof steps}
\end{figure}

\begin{itemize}
\item[Step 1] $ t_{\ee_j,\mathbf 0} = \bar t_{\ee_j,\mathbf 0}$ for $j>2K$;
\item[Step 2] $ t_{\ee_j,\ee_k} = \bar t_{\ee_j,\ee_k}$ for $j>2K$ and $k\in\{1,\cdots,K\}$;
\item[Step 3] $ t_{\ee_j,\mathbf 0} = \bar t_{\ee_j,\mathbf 0}$, $ t_{\ee_j,\ee_k} = \bar t_{\ee_j,\ee_k}$,
$p_{\mathbf 0} = \bar p_{\mathbf 0}$ 
 and 
$p_{\ee_k} = \bar p_{\ee_k}$  for $j\in\{1,\cdots,2K\}$ and $k\in\{1,\cdots,K\}$;
\item[Step 4] $ t_{\ee_j,\ee_{h_1}+\ee_{h_2}} = \bar t_{\ee_j,\ee_{h_1}+\ee_{h_2}}$ and 
$p_{\ee_{h_1}+\ee_{h_2}} = \bar p_{\ee_{h_1}+\ee_{h_2}}$
 for $j\in\{1,\cdots,J\}$ and  $1\leq h_1\neq h_2\leq K$;
\item[Step 5] 
$t_{\ee_j,\sum_{i=1}^{k}\ee_{h_i}}=\bar t_{\ee_j,\sum_{i=1}^{k}\ee_{h_i}}$ and  
$p_{\sum_{i=1}^{k}\ee_{h_i}}=\bar p_{\sum_{i=1}^{k}\ee_{h_i}}$
for $j\in\{1,\cdots,J\}$ and  $1\leq h_1\neq \cdots\neq h_k\leq K$ with any $2<k\leq K$.
\end{itemize}
For a better illustration, Figure 1 specifies the corresponding components of the $\Theta$ matrix and the $\pp$ vector that we will focus on in each step.
Combining the results in the five steps, we have the desired conclusion that 
$\Theta =\bar \Theta \mbox{ and } \pp=\bar\pp.$

In order to establish Steps 1--5, we need to incorporate into \eqref{mleeq} the constraints of the parameters under the restricted latent class models. 
This is achieved by the following linear transformation of the $T$-matrix in Proposition \ref{trans}.
We  extend the definition of $T$-matrix through \eqref{defT} to include $\Theta \not\in[0,1]^{J\times 2^K}$, where $t_{\rr,\aalpha}(Q,\Theta)$ will no longer correspond to probabilities. 
We  order the column indices of the $T$-matrix from left to right as $(\mathbf 0, \ee_1,\cdots, \ee_K, \ee_1+\ee_2,\cdots,\ee_{K-1}+\ee_{K}, \cdots, \mathbf 1)$ and the row indices from top to bottom as $(\mathbf 0, \ee_1,\cdots, \ee_J, \ee_1+\ee_2,\cdots,\ee_{J-1}+\ee_{J}, \cdots, \mathbf 1)$.

\begin{proposition}\label{trans}
For any $\ttt^*=(\theta_1^*,...,\theta_J^*)^\top \in \mathbb{R}^J$, there exists an invertible
matrix $D(\ttt^*)$ depending solely on $\ttt^*$, such that the matrix $D(\ttt^*)$ is lower triangular with diagonal $\text{diag}\{D(\ttt^*)\} = \bfone$, and 
$$T(Q,\Theta-\ttt^*\mathbf 1^\top) = D(\ttt^*) T(Q,\Theta).$$
\end{proposition}
\noindent Proposition \ref{trans} shows that 
equation \eqref{mleeq}
is equivalent to $$T(Q,\Theta-\ttt^*\mathbf 1^\top) \pp 
= T(Q,\bar\Theta-\ttt^*\mathbf 1^\top)\bar \pp.$$
Note that the vector product 
$\ttt^*\mathbf 1^\top$ is a $J\times 2^K$ matrix with the $j$th row equal to 
$\theta_j^*\mathbf 1^\top = (\theta_j^*, \cdots, \theta_j^*)_{1\times 2^K}$, and
 the  $j$th row vector of $\Theta-\ttt^*\mathbf 1^\top$ is $(\theta_{j,\aaa} - \theta_j^*; ~\aaa\in\{0,1\}^K ).$ Thus, if we take $\theta_j^*$ equal to $\theta_{j,\aaa}$,  the corresponding element in $\Theta-\ttt^*\mathbf 1^\top$ will become $0$. By properly choosing the vector $\ttt^*$ according to the $Q$-restrictions, we can then make certain elements in $T(Q,\Theta-\ttt^*\mathbf 1^\top)$ to be $0$.
For instance, if we choose $\theta^*_1= t_{\ee_1,\mathbf 1}(Q,\Theta)$, then we have the transformed matrix elements
$t_{\ee_1,\aaa}(Q,\Theta-\ttt^*\mathbf 1^\top) = 0$  for all $\aaa\succeq \qq_1$.
 This nice algebraic structure makes the transformed $T$-matrix much easier to work with and plays a key role in the following proof.
 
\paragraph{Step 1} 
We apply the result in Proposition \ref{trans}.
Define
$$\ttt^*=\big(
~\underbrace{\bar t_{\ee_1,\mathbf 1},\cdots,\bar t_{\ee_K,\mathbf 1}}_{K},~
 \underbrace{t_{\ee_{K+1},\mathbf 1},\cdots, t_{\ee_{2K},\mathbf 1}}_{K}, ~
\underbrace{0,\cdots,0}_{J-2K}~\big)^\top,
$$
  and \eqref{mleeq} gives
\begin{equation}\label{Apr14_1}
 T(Q, \Theta-\ttt^*\mathbf 1^\top)\pp= T(Q,\bar \Theta-\ttt^*\mathbf 1^\top)\bar\pp.
 \end{equation}
Note that for any $k\in\{1,\cdots,K\}$, 
$\bar t_{\ee_k,\aaa} - \theta^*_k = \bar t_{\ee_k,\aaa} - \bar t_{\ee_k,\mathbf 1}=0$ if $\aaa\succeq \ee_k$, and similarly, 
$ t_{\ee_{K+k},\aaa} - \theta^*_{K+k} =  t_{\ee_{K+k},\aaa} -  t_{\ee_{K+k},\mathbf 1}=0$ if $\aaa\succeq \ee_k$.

Consider the row vector of  $T(Q, \Theta-\ttt^*\mathbf 1^\top)$ corresponding to 
$\rr =\sum_{k=1}^{2K}\ee_k$, i.e., $T_{\sum_{k=1}^{2K}\ee_k,\Cdot}(Q, \Theta-\ttt^*\mathbf 1^\top)$.
From the definition form \eqref{defT} of the $T$-matrix, we know  
\begin{eqnarray*}
T_{\sum_{k=1}^{2K}\ee_k,\Cdot}(Q, \Theta-\ttt^*\mathbf 1^\top)
&=& \bigodot_{k=1}^{2K}\left\{T_{\ee_k,\Cdot}(Q, \Theta-\ttt^*\mathbf 1^\top)\right\}\\
&=&\left(\prod_{k=1}^K( t_{\ee_k,\mathbf 0}-\bar t_{\ee_k,\mathbf 1})\prod_{k=K+1}^{2K}( t_{\ee_k,\mathbf 0}- t_{\ee_k,\mathbf 1}),~~\mathbf 0^\top\right).
\end{eqnarray*}
That is, the last $2^K-1$ elements  of the row vector $T_{\sum_{k=1}^{2K}\ee_k,\Cdot}(Q, \Theta-\ttt^*\mathbf 1^\top)$  are 0.
Next we show that the first element  of $T_{\sum_{k=1}^{2K}\ee_k,\Cdot}(Q, \Theta-\ttt^*\mathbf 1^\top)$ is nonzero, i.e.,
$$\prod_{k=1}^K(t_{\ee_k,\mathbf 0}-\bar t_{\ee_k,\mathbf 1})
\prod_{k=K+1}^{2K}(t_{\ee_k,\mathbf 0}- t_{\ee_k,\mathbf 1})\neq 0.$$
We introduce the following lemma, whose proof is in Section 4.2.
\begin{lemma}\label{lemma0}
Under the conditions of Theorem 1, if \eqref{mleeq} holds, then for any $1\leq k\leq K$ and $\aaa^*\succeq\ee_k$
$$t_{\ee_k,\mathbf 0}\neq \bar t_{\ee_k,\aaa^*}, 
~
t_{\ee_k,\aaa^*}\neq \bar t_{\ee_k,\mathbf 0},~
t_{\ee_{K+k},\mathbf 0}\neq \bar t_{\ee_{K+k},\aaa^*} 
\mbox{ and }
t_{\ee_{K+k},\aaa^*}\neq \bar t_{\ee_{K+k},\mathbf 0}.
$$
\end{lemma}

\noindent 
Lemma \ref{lemma0} implies that $\prod_{k=1}^K(t_{\ee_k,\mathbf 0}-\bar t_{\ee_k,\mathbf 1})\neq 0.$
In addition,  from the assumption that $t_{\ee_k,\mathbf 0}< t_{\ee_k,\mathbf 1}$ for  $k\in \{1,\ldots,2K\}$, we have 
$\prod_{k=K+1}^{2K}(t_{\ee_k,\mathbf 0}- t_{\ee_k,\mathbf 1})\neq 0$. 
Thus the first element of the row vector $T_{\sum_{k=1}^{2K}\ee_k,\Cdot}(Q, \Theta-\ttt^*\mathbf 1^\top)$ is not 0.


Similarly, by doing the same transformation, we have
$$T_{\sum_{k=1}^{2K}\ee_k,\Cdot}(Q,\bar \Theta-\ttt^*\mathbf 1^\top)=\left(\prod_{k=1}^K(\bar t_{\ee_k,\mathbf 0}-\bar t_{\ee_k,\mathbf 1})\prod_{k=K+1}^{2K}(\bar t_{\ee_k,\mathbf 0}- t_{\ee_k,\mathbf 1}),~\mathbf 0^\top\right),$$
where the first element 
 $\prod_{k=1}^K(\bar t_{\ee_k,\mathbf 0}-\bar t_{\ee_k,\mathbf 1})\prod_{k=K+1}^{2K}(\bar t_{\ee_k,\mathbf 0}- t_{\ee_k,\mathbf 1})\neq 0$
 and the rest elements are 0.

Now consider any $j> 2K$. The row vector of  $T(Q, \Theta-\ttt^*\mathbf 1^\top)$ corresponding to $\rr= \ee_j+\sum_{k=1}^{2K}\ee_k$ equals 
\begin{eqnarray*}
&& T_{\ee_j+\sum_{k=1}^{2K}\ee_k,\Cdot}(Q, \Theta-\ttt^*\mathbf 1^\top)\\
&=&T_{\ee_j,\Cdot}(Q, \Theta-\ttt^*\mathbf 1^\top) \odot \left[\bigodot_{k=1}^{2K}\left\{T_{\ee_k,\Cdot}(Q, \Theta-\ttt^*\mathbf 1^\top)\right\}\right]\\
&=&
\left( t_{\ee_j,\mathbf 0}\times
\prod_{k=1}^K( t_{\ee_k,\mathbf 0}-\bar t_{\ee_k,\mathbf 1})
\prod_{k=K+1}^{2K}( t_{\ee_k,\mathbf 0}- t_{\ee_k,\mathbf 1}),\quad
\mathbf 0^\top\right)\\
&=&
 t_{\ee_j,\mathbf 0}\cdot T_{\sum_{k=1}^{2K}\ee_k,\Cdot}(Q, \Theta-\ttt^*\mathbf 1^\top)
\end{eqnarray*}
and similarly 
\begin{eqnarray*}
&& T_{\ee_j+\sum_{k=1}^{2K}\ee_k,\Cdot}(Q,\bar \Theta-\ttt^*\mathbf 1^\top)\\
&=&
\left(\bar t_{\ee_j,\mathbf 0}\times
\prod_{k=1}^K(\bar t_{\ee_k,\mathbf 0}-\bar t_{\ee_k,\mathbf 1})
\prod_{k=K+1}^{2K}(\bar t_{\ee_k,\mathbf 0}- t_{\ee_k,\mathbf 1}),
\mathbf 0^\top\right)\\
&=& \bar t_{\ee_j,\mathbf 0}\cdot T_{\sum_{k=1}^{2K}\ee_k,\Cdot}(Q, \bar\Theta-\ttt^*\mathbf 1^\top).
\end{eqnarray*}
By equation \eqref{Apr14_1}
\begin{eqnarray*}
T_{\ee_j+\sum_{k=1}^{2K}\ee_k,\Cdot}(Q, \Theta-\ttt^*\mathbf 1^\top)\pp
&=& 
T_{\ee_j+\sum_{k=1}^{2K}\ee_k,\Cdot}(Q,\bar \Theta-\ttt^*\mathbf 1^\top)\bar\pp
\\
\mbox{and }\quad
T_{\sum_{k=1}^{2K}\ee_k,\Cdot}(Q, \Theta-\ttt^*\mathbf 1^\top)\pp
&=&
T_{\sum_{k=1}^{2K}\ee_k,\Cdot}(Q,\bar \Theta-\ttt^*\mathbf 1^\top)\bar\pp.
\end{eqnarray*}
Thus for any $j>2K$,
$$ t_{\ee_j,\mathbf 0} =
\frac{T_{\ee_j+\sum_{k=1}^{2K}\ee_k,\Cdot}(Q, \Theta-\ttt^*\mathbf 1^\top)\pp}
{T_{\sum_{k=1}^{2K}\ee_k,\Cdot}(Q, \Theta-\ttt^*\mathbf 1^\top)\pp}
=
\frac{T_{\ee_j+\sum_{k=1}^{2K}\ee_k,\Cdot}(Q,\bar \Theta-\ttt^*\mathbf 1^\top)\bar\pp}
{T_{\sum_{k=1}^{2K}\ee_k,\Cdot}(Q,\bar \Theta-\ttt^*\mathbf 1^\top)\bar\pp}
= \bar t_{\ee_j,\mathbf 0}.$$
This completes Step 1.

\paragraph{Step 2} 
To better illustrate our idea, we first focus on the column with respect to $\aaa=\ee_1$ and show $$t_{\ee_j,\ee_1} =\bar t_{\ee_j,\ee_1} 
\mbox{ for $j>2K$.} $$
We redefine the $\ttt^*$ vector as 
$$\ttt^*=\big(~
\underbrace{\bar t_{\ee_1,\mathbf 0},\bar t_{\ee_2,\mathbf 1},\cdots,\bar t_{\ee_K,\mathbf 1}}_K,~
\underbrace{ t_{\ee_{K+1},\mathbf 0},  t_{\ee_{K+2},\mathbf 1},
\cdots, t_{\ee_{2K},\mathbf 1}}_K, ~
\underbrace{0,\cdots,0}_{J-2K}~\big)^\top,$$
where the first element is $\bar t_{\ee_1,\mathbf 0}$ and the $(K+1)$th element is $ t_{\ee_{K+1},\mathbf 0}$ while the other elements are the same as the $\ttt^*$ vector  taken in Step 1. 
For the chosen $\ttt^*$, the row vectors of the transformed $T$-matrices corresponding to items 1,..., $2K$, i.e., $\rr=\sum_{k=1}^{2K}\ee_k$,  are
\begin{eqnarray}\label{Apr15_2}
&&T_{\sum_{k=1}^{2K}\ee_k,\Cdot}(Q, \Theta-\ttt^*\mathbf 1^\top)
= \bigodot_{k=1}^{2K}\left\{T_{\ee_k,\Cdot}(Q, \Theta-\ttt^*\mathbf 1^\top)\right\}
\notag\\
&=&\biggr(0,~ ( t_{\ee_1,\ee_1}-\bar t_{\ee_1,\mathbf 0})
\prod_{k=2}^K( t_{\ee_k,\ee_1}-\bar t_{\ee_k,\mathbf 1})
\notag\\
&&\quad\quad \times
( t_{\ee_{K+1},\ee_{1}}- t_{\ee_{K+1},\mathbf 0})
\prod_{k=K+2}^{2K}( t_{\ee_k,\ee_{1}}- t_{\ee_k,\mathbf 1}),
~\mathbf 0^\top\biggr),
\end{eqnarray}
and 
\begin{eqnarray}\label{Apr15_3}
&&T_{\sum_{k=1}^{2K}\ee_k,\Cdot}(Q, \bar\Theta-\ttt^*\mathbf 1^\top)
= \bigodot_{k=1}^{2K}\left\{T_{\ee_k,\Cdot}(Q, \bar\Theta-\ttt^*\mathbf 1^\top)\right\}
\notag\\
&=&\biggr(0,~ (\bar t_{\ee_1,\ee_1}-\bar t_{\ee_1,\mathbf 0})
\prod_{k=2}^K(\bar t_{\ee_k,\ee_1}-\bar t_{\ee_k,\mathbf 1})
\notag\\
&&\quad\quad\times
(\bar t_{\ee_{K+1},\ee_{1}}- t_{\ee_{K+1},\mathbf 0})
\prod_{k=K+2}^{2K}(\bar t_{\ee_k,\ee_{1}}- t_{\ee_k,\mathbf 1}),~\mathbf 0^\top\biggr).
\end{eqnarray}
We now show the second elements of the above two vectors are nonzero. 
We need the following lemma, which is proved in Section 4.2.
\begin{lemma}\label{lemma1}
Under the conditions of Theorem 1, if \eqref{mleeq} holds, then for any $1\leq k\neq h\leq K$,
$$t_{\ee_k, \ee_h}\neq \bar t_{\ee_k, \mathbf 1},  ~~
t_{\ee_k, \mathbf 1}\neq \bar t_{\ee_k, \ee_h},~~
t_{\ee_{K+k}, \ee_h}\neq \bar t_{\ee_{K+k}, \mathbf 1}, \mbox{ and }
~t_{\ee_{K+k}, \mathbf 1}\neq \bar t_{\ee_{K+k}, \ee_h}.$$
\end{lemma}
{Consider vector \eqref{Apr15_2}. 
Lemma \ref{lemma0} implies that   
$( t_{\ee_1,\ee_1}-\bar t_{\ee_1,\mathbf 0})\neq 0$, and 
 Lemma \ref{lemma1} implies  
$$\prod_{k=2}^K( t_{\ee_k,\ee_1}-\bar t_{\ee_k,\mathbf 1})\neq 0.$$
Moreover, for the term  $( t_{\ee_{K+1},\ee_{1}}- t_{\ee_{K+1},\mathbf 0})$, since the $(K+1)$th item only requires the first attribute, i.e., the $\qq$-vector is $\ee_1$, we know
$t_{\ee_{K+1},\ee_{1}} =  t_{\ee_{K+1},\mathbf 1} > t_{\ee_{K+1},\mathbf 0}.$
Similarly, we have 
$$\prod_{k=K+2}^{2K}( t_{\ee_k,\ee_{1}}- t_{\ee_k,\mathbf 1})\neq 0.$$
The above results implies that the second element of \eqref{Apr15_2} is nonzero. From a similar argument, the second element  of  \eqref{Apr15_3} is also nonzero. 
}

Now consider any $j\geq 2K+1$. 
We have  
\begin{eqnarray*}
&&T_{\ee_j+\sum_{k=1}^{2K}\ee_k,\Cdot}(Q, \Theta-\ttt^*\mathbf 1^\top)\\
&=&\biggr(0,~ t_{\ee_j,\ee_1}( t_{\ee_1,\ee_1}-\bar t_{\ee_1,\mathbf 0})
\prod_{k=2}^K( t_{\ee_k,\ee_1}-\bar t_{\ee_k,\mathbf 1})\\
&&\quad\quad \times
( t_{\ee_{K+1},\ee_{1}}- t_{\ee_{K+1},\mathbf 0})
\prod_{k=K+2}^{2K}( t_{\ee_k,\ee_{1}}- t_{\ee_k,\mathbf 1}),
~\mathbf 0^\top\biggr)\\
&=& t_{\ee_j,\ee_1}\cdot T_{\sum_{k=1}^{2K}\ee_k,\Cdot}(Q, \Theta-\ttt^*\mathbf 1^\top) 
\end{eqnarray*}
and 
\begin{eqnarray*}
&&T_{\ee_j+\sum_{k=1}^{2K}\ee_k,\Cdot}(Q, \bar\Theta-\ttt^*\mathbf 1^\top)\\
&=&\biggr(0,~  \bar t_{\ee_j,\ee_1}(\bar t_{\ee_1,\ee_1}-\bar t_{\ee_1,\mathbf 0})
\prod_{k=2}^K(\bar t_{\ee_k,\ee_1}-\bar t_{\ee_k,\mathbf 1})\\
&&\quad\quad\times
(\bar t_{\ee_{K+1},\ee_{1}}- t_{\ee_{K+1},\mathbf 0})
\prod_{k=K+2}^{2K}(\bar t_{\ee_k,\ee_{1}}- t_{\ee_k,\mathbf 1}),~\mathbf 0^\top\biggr)\\
&=&\bar t_{\ee_j,\ee_1}\cdot 
T_{\sum_{k=1}^{2K}\ee_k,\Cdot}(Q, \bar\Theta-\ttt^*\mathbf 1^\top) .
\end{eqnarray*}
As in Step 1, since 
\begin{eqnarray*}
T_{\ee_j+\sum_{k=1}^{2K}\ee_k,\Cdot}(Q, \Theta-\ttt^*\mathbf 1^\top)\pp
&=&T_{\ee_j+\sum_{k=1}^{2K}\ee_k,\Cdot}(Q,\bar \Theta-\ttt^*\mathbf 1^\top)\bar\pp,\\
T_{\sum_{k=1}^{2K}\ee_k,\Cdot}(Q, \Theta-\ttt^*\mathbf 1^\top)\pp
&=&T_{\sum_{k=1}^{2K}\ee_k,\Cdot}(Q,\bar \Theta-\ttt^*\mathbf 1^\top)\bar\pp,
\end{eqnarray*}
we have 
$$t_{\ee_j,\ee_1} 
=
\frac{T_{\ee_j+\sum_{k=1}^{2K}\ee_k,\Cdot}(Q, \Theta-\ttt^*\mathbf 1^\top)\pp}
{T_{\sum_{k=1}^{2K}\ee_k,\Cdot}(Q, \Theta-\ttt^*\mathbf 1^\top)\pp}
=
\frac{T_{\ee_j+\sum_{k=1}^{2K}\ee_k,\Cdot}(Q,\bar \Theta-\ttt^*\mathbf 1^\top)\bar\pp}
{T_{\sum_{k=1}^{2K}\ee_k,\Cdot}(Q,\bar \Theta-\ttt^*\mathbf 1^\top)\bar\pp}
= \bar t_{\ee_j,\ee_1}.$$

The above  argument can be easily generalized to any $1<h\leq K$. Redefine  
\begin{eqnarray*}
\ttt^* &=& \big(~\underbrace{\bar t_{\ee_1,\mathbf 1},\cdots, \bar t_{\ee_{h-1},\mathbf 1},\bar t_{\ee_{h},\mathbf 0},\bar t_{\ee_{h+1},\mathbf 1},\cdots,\bar t_{\ee_K,\mathbf 1}}_K,\\
&& \underbrace{t_{\ee_{K+1},\mathbf 1},\cdots,  t_{\ee_{K+h-1},\mathbf 1}, t_{\ee_{K+h},\mathbf 0}, t_{\ee_{K+h+1},\mathbf 1},\cdots, t_{\ee_{2K},\mathbf 1}}_K,
~
\underbrace{0,\cdots,0}_{J-2K}~\big)^\top.
\end{eqnarray*}
Following a similar argument as above, we can get  for any $j\geq 2K+1$ and $k\in\{1,\cdots, K\}$, 
$t_{\ee_j,\ee_k} = \bar t_{\ee_j,\ee_k}.$
This completes Step 2. 

\paragraph{Step 3}
From  assumption \ref{C2},  for any $1\leq k\leq K$,  
$$(t_{\ee_{2K+1},\ee_k}, \cdots, t_{\ee_{J},\ee_k})^\top \neq (t_{\ee_{2K+1},\mathbf 0}, \cdots, t_{\ee_{J},\mathbf 0})^\top.$$ 
Then $(1, t_{\ee_{2K+1},\ee_k}, \cdots, t_{\ee_{J},\ee_k})^\top$ and $\allowbreak(1, t_{\ee_{2K+1},\mathbf 0}, \cdots, t_{\ee_{J},\mathbf 0})^\top$ are not proportional to each other.
There exists a $(J-2K+1)$-dimensional row vector $\uu_k$ such that 
$$b_k:= \uu_k(1, t_{\ee_{2K+1},\ee_k}, \cdots, t_{\ee_{J},\ee_k})^\top \neq0 \mbox{ and }
 \uu_k(1, t_{\ee_{2K+1},\mathbf 0}, \cdots, t_{\ee_{J},\mathbf 0})^\top =0.$$
Consider matrix 
$$A(Q, \Theta) 
= \left(\begin{array}{c}
\mathbf 1^\top\\
T_{\ee_{2K+1},\Cdot}(Q,\Theta)\\
T_{\ee_{2K+2},\Cdot}(Q,\Theta)\\
\vdots\\
T_{\ee_J,\Cdot}(Q,\Theta)
\end{array}\right).$$
From the first two steps, we know that the first $K+1$ columns of $A(Q,\Theta)$ and $A(Q,\bar\Theta)$  are equal.
For simplicity, we write $A(Q,\Theta)$ and $A(Q,\bar\Theta)$ as $A$ and $\bar A$, respectively.
 Then we have 
 \begin{eqnarray}\label{ua}
 \uu_k A &=&  (0, {*,\ldots,*}, \underbrace{b_k}_{column ~\ee_k}, ~{*,\ldots,*}),
 \\
  \uu_k \bar A &=&  (0, {*,\ldots,*}, \underbrace{b_k}_{column ~\ee_k}, ~{*,\ldots,*}),
  \notag
 \end{eqnarray}
 where $*$'s are unspecified values. 

We use the above results to prove Step 3.
For $h\in \{1,\cdots,K\}$, redefine  
\begin{eqnarray*}
\ttt^* &=&
\big(
\underbrace{\bar t_{\ee_1,\mathbf 1},\cdots, \bar t_{\ee_{h-1},\mathbf 1},0,\bar t_{\ee_{h+1},\mathbf 1},\cdots,\bar t_{\ee_K,\mathbf 1}}_K,\\
&& 
\underbrace{t_{\ee_{K+1},\mathbf 1},\cdots, t_{\ee_{K+h-1},\mathbf 1}, 0,
 t_{\ee_{K+h+1},\mathbf 1},\cdots, t_{\ee_{2K},\mathbf 1}}_{K},~
\underbrace{0,\cdots,0}_{J-2K}~\big)^\top
\end{eqnarray*}
and we have $ T(Q, \Theta-\ttt^*\mathbf 1^\top)\pp= T(Q,\bar \Theta-\ttt^*\mathbf 1^\top)\bar\pp$.
{With such a choice of $\ttt^*$,  for any $k\in\{1,иии,K\}$ and $k\neq h$,
$\bar t_{\ee_k,\aaa} - \theta^*_k = \bar t_{\ee_k,\aaa} - \bar t_{\ee_k,\mathbf 1}=0$ if $\aaa\succeq \ee_k$, and similarly, 
$ t_{\ee_{K+k},\aaa} - \theta^*_{K+k} =  t_{\ee_{K+k},\aaa} -  t_{\ee_{K+k},\mathbf 1}=0$ if $\aaa\succeq \ee_k$.
}

Consider the row vectors of $T$-matrices corresponding to items 1,..., $2K$ except $h$ and $K+h$, i.e., $\rr=\sum_{k=1}^{2K}\ee_k -\ee_h-\ee_{K+h}$.
We have 
\begin{eqnarray}\label{Apr14_2}
&& T_{\sum_{k=1}^{2K}\ee_k -\ee_h-\ee_{K+h},\Cdot}(Q, \Theta-\ttt^*\mathbf 1^\top)\notag\\
&=&\biggr( 
\prod_{\substack{k=1,\cdots, K,\\ k\neq h}}
( t_{\ee_k,\mathbf 0}-\bar t_{\ee_k,\mathbf 1})\times
\prod_{\substack{k=K+1,\cdots, 2K, \\ k\neq K+h}}
( t_{\ee_k,\mathbf 0}- t_{\ee_k,\mathbf 1}),~\mathbf 0^\top,\notag\\
&& \underbrace{
\prod_{\substack{k=1,\cdots, K,\\ k\neq h}}
( t_{\ee_k,\ee_h}-\bar t_{\ee_k,\mathbf 1})\times
\prod_{\substack{k=K+1,\cdots, 2K,\\ k\neq K+h}}
( t_{\ee_k,\ee_h}- t_{\ee_k,\mathbf 1})}
_{column~\ee_h},~\mathbf 0^\top\biggr),
\end{eqnarray}
where the second product term corresponds to column $\ee_h$,
and
\begin{eqnarray}\label{Apr14_3}
&&
T_{\sum_{k=1}^{2K}\ee_k -\ee_h-\ee_{K+h},\Cdot}(Q, \bar\Theta-\ttt^*\mathbf 1^\top)\notag\\
&=&\biggr(
\prod_{\substack{k=1,\cdots, K,\\ k\neq h}}
(\bar t_{\ee_k,\mathbf 0}-\bar t_{\ee_k,\mathbf 1})\times
\prod_{\substack{k=K+1,\cdots, 2K,\\ k\neq K+h}}
( \bar t_{\ee_k,\mathbf 0}- t_{\ee_k,\mathbf 1}),~\mathbf 0^\top\notag\\
&& \underbrace{
\prod_{\substack{k=1,\cdots, K,\\ k\neq h}}
(\bar t_{\ee_k,\ee_h}-\bar t_{\ee_k,\mathbf 1})\times
\prod_{\substack{k=K+1,\cdots, 2K, \\ k\neq K+h}}
(\bar t_{\ee_k,\ee_h}- t_{\ee_k,\mathbf 1})}_{column~\ee_h}
,~\mathbf 0^\top\biggr).
\end{eqnarray}
From Lemmas \ref{lemma0}--\ref{lemma1} and the model assumption, we know the product components in \eqref{Apr14_2} and \eqref{Apr14_3} are nonzero.
Adding item $h$ into the above combinations, the row vectors corresponding to $\rr=\sum_{k=1}^{2K}\ee_k -\ee_{K+h}$ equal to
\begin{eqnarray}\label{Apr14_4}
&& T_{\sum_{k=1}^{2K}\ee_k -\ee_{K+h},\Cdot}(Q, \Theta-\ttt^*\mathbf 1^\top)
\notag\\
&=&\biggr( t_{\ee_h,\mathbf 0}\times 
\prod_{\substack{k=1,\cdots, K,\\ k\neq h}}
( t_{\ee_k,\mathbf 0}-\bar t_{\ee_k,\mathbf 1})\times
\prod_{\substack{k=K+1,\cdots, 2K, \\ k\neq K+h}}
( t_{\ee_k,\mathbf 0}- t_{\ee_k,\mathbf 1}),~\mathbf 0^\top
\notag\\
&& \underbrace{
t_{\ee_h, \ee_h}\times 
\prod_{\substack{k=1,\cdots, K,\\ k\neq h}}
( t_{\ee_k,\ee_h}-\bar t_{\ee_k,\mathbf 1})\times
\prod_{\substack{k=K+1,\cdots, 2K, \\ k\neq K+h}}
( t_{\ee_k,\ee_h}- t_{\ee_k,\mathbf 1})}
_{column~\ee_h},~\mathbf 0^\top\biggr),
\end{eqnarray}
and
\begin{eqnarray}\label{Apr14_5}
&&
T_{\sum_{k=1}^{2K}\ee_k -\ee_{K+h},\Cdot}(Q, \bar\Theta-\ttt^*\mathbf 1^\top)
\notag\\
&=&\biggr( \bar t_{\ee_h,\mathbf 0}\times 
\prod_{\substack{k=1,\cdots, K,\\ k\neq h}}
(\bar t_{\ee_k,\mathbf 0}-\bar t_{\ee_k,\mathbf 1})\times
\prod_{\substack{k=K+1,\cdots, 2K,\\ k\neq K+h}}
( \bar t_{\ee_k,\mathbf 0}- t_{\ee_k,\mathbf 1}),~\mathbf 0^\top
\notag\\
&& \underbrace{
\bar t_{\ee_h, \ee_h}\times
\prod_{\substack{k=1,\cdots, K,\\ k\neq h}}
(\bar t_{\ee_k,\ee_h}-\bar t_{\ee_k,\mathbf 1})\times
\prod_{\substack{k=K+1,\cdots, 2K,\\ k\neq K+h}}
(\bar t_{\ee_k,\ee_h}- t_{\ee_k,\mathbf 1})}_{column~\ee_h}
,~\mathbf 0^\top\biggr).
\end{eqnarray}

Take the element-wise product of the row vectors: $\uu_h A$ defined in \eqref{ua} and the vector in \eqref{Apr14_2}.
We have
\begin{eqnarray*}
&&(\uu_hA)\odot T_{\sum_{k=1}^{2K}\ee_k -\ee_h-\ee_{K+h},\Cdot}(Q, \Theta-\ttt^*\mathbf 1^\top)\\
&=&\biggl(\mathbf 0, ~\underbrace{b_h\prod_{\substack{ k=1,\cdots, K,\\ k\neq h}}( t_{\ee_k,\ee_h}-\bar t_{\ee_k,\mathbf 1})\times
\prod_{\substack{k=K+1,\cdots, 2K, \\ k\neq K+h}}( t_{\ee_k,\ee_h}- t_{\ee_k,\mathbf 1})}_{column~\ee_h},~\mathbf 0^\top\biggr).
\end{eqnarray*}
From $\uu_h\bar A$ in \eqref{ua} and the vector in \eqref{Apr14_3}
\begin{eqnarray*}
&&(\uu_h\bar A)\odot T_{\sum_{k=1}^{2K}\ee_k -\ee_h-\ee_{K+h},\Cdot}(Q, \bar\Theta-\ttt^*\mathbf 1^\top)\\
&=&\biggl(\mathbf 0, ~\underbrace{b_h\prod_{\substack{ k=1,\cdots, K,\\ k\neq h}}(\bar t_{\ee_k,\ee_h}-\bar t_{\ee_k,\mathbf 1})\times
\prod_{\substack {k=K+1,\cdots, 2K,\\ k\neq K+h}}
(\bar t_{\ee_k,\ee_h}- t_{\ee_k,\mathbf 1})}_{column~\ee_h},~\mathbf 0^\top\biggr),
\end{eqnarray*}
Similarly,  the element-wise product of $\uu_h A$ and \eqref{Apr14_4}
gives 
\begin{eqnarray}\label{gx1}
&&(\uu_hA)\odot T_{\sum_{k=1}^{2K}\ee_k-\ee_{K+h},\Cdot}(Q, \Theta-\ttt^*\mathbf 1^\top)\notag\\
&=&\biggr(\mathbf 0, ~\underbrace{b_h t_{\ee_h,\ee_h}\prod_{\substack{ k=1,\cdots, K,\\ k\neq h}}( t_{\ee_k,\ee_h}-\bar t_{\ee_k,\mathbf 1})\times
\prod_{\substack{k=K+1,\cdots, 2K, \\k\neq K+h}}
( t_{\ee_k,\ee_h}- t_{\ee_k,\mathbf 1})}_{column~\ee_h},~\mathbf 0^\top\biggr)\notag\\
&=& t_{\ee_h,\ee_h} \cdot \left\{(\uu_hA)\odot T_{\sum_{k=1}^{2K}\ee_k -\ee_h-\ee_{K+h},\Cdot}(Q, \Theta-\ttt^*\mathbf 1^\top)\right\},\notag\\
\end{eqnarray}
and the element-wise product of $\uu_h \bar A$  and \eqref{Apr14_5} gives
\begin{eqnarray}\label{gx2}
&&(\uu_h\bar A)\odot
T_{\sum_{k=1}^{2K}\ee_k-\ee_{K+h},\Cdot}(Q, \bar\Theta-\ttt^*\mathbf 1^\top)\notag\\
&=&\biggr(\mathbf 0, ~\underbrace{b_h\bar t_{\ee_h,\ee_h}
\prod_{\substack{k=1,\cdots, K,\\ k\neq h}}
(\bar t_{\ee_k,\ee_h}-\bar t_{\ee_k,\mathbf 1})\times
\prod_{\substack{k=K+1,\cdots, 2K,\\ k\neq K+h}}
(\bar t_{\ee_k,\ee_h}- t_{\ee_k,\mathbf 1})}_{column~\ee_h},~\mathbf 0^\top\biggr)\notag\\
&=&\bar t_{\ee_h,\ee_h} \cdot \left\{(\uu_h\bar A)\odot T_{\sum_{k=1}^{2K}\ee_k -\ee_h-\ee_{K+h},\Cdot}(Q, \bar\Theta-\ttt^*\mathbf 1^\top)\right\}.\notag\\
\end{eqnarray}

 From the equation that
$ T(Q, \Theta-\ttt^*\mathbf 1^\top)\pp= T(Q,\bar \Theta-\ttt^*\mathbf 1^\top)\bar\pp,$
we know 
\begin{eqnarray*}
&& \left\{(\uu_hA)\odot T_{\sum_{k=1}^{2K}\ee_k -\ee_h-\ee_{K+h},\Cdot}(Q, \Theta-\ttt^*\mathbf 1^\top)\right\}\pp\\
&=&
\left\{(\uu_h\bar A)\odot T_{\sum_{k=1}^{2K}\ee_k -\ee_h-\ee_{K+h},\Cdot}(Q, \bar\Theta-\ttt^*\mathbf 1^\top)\right\}\bar\pp
\end{eqnarray*}
and
\begin{eqnarray*}
&& \left\{(\uu_hA)\odot T_{\sum_{k=1}^{2K}\ee_k -\ee_{K+h},\Cdot}(Q, \Theta-\ttt^*\mathbf 1^\top)\right\}\pp\\
&=&
\left\{(\uu_h\bar A)\odot T_{\sum_{k=1}^{2K}\ee_k -\ee_{K+h},\Cdot}(Q, \bar\Theta-\ttt^*\mathbf 1^\top)\right\}\bar\pp.
\end{eqnarray*}
Therefore,  \eqref{gx1} and \eqref{gx2} imply that  for $h=1,\cdots, K$,
\begin{equation}\label{App}
t_{\ee_h,\ee_h} = \bar t_{\ee_h,\ee_h}. 
\end{equation}
Similarly, we have $t_{\ee_{K+h},\ee_{h}} = \bar t_{\ee_{K+h},\ee_{h}}$.

Furthermore,  there exists row vector $\vv_k$ such that 
$$ \vv_k(1, t_{\ee_{2K+1},\ee_k}, \cdots, t_{\ee_{J},\ee_k})^\top =0 ~\mbox{ and }~
 \vv_k(1,  t_{\ee_{2K+1},\mathbf 0}, \cdots, t_{\ee_{J},\mathbf 0})^\top \neq 0.$$
A similar argument then gives 
$$t_{\ee_h,\mathbf 0} = \bar t_{\ee_h,\mathbf 0} \mbox{ for } h=1,\cdots, 2K.$$

Before to prove $t_{\ee_j,\ee_h} = \bar t_{\ee_j,\ee_h}$
for the rest $j\in\{1,\cdots, 2K\}$ and $h\in\{1,\cdots, K\}$,
we first  show $p_{\bfzero} = \bar p_{\bfzero}$ and  $p_{\ee_h} = \bar p_{\ee_h}$ 
for $h\in\{1,\cdots, K\}$. 
Take
\begin{eqnarray*}
\ttt^* = \big(~\underbrace{ t_{\ee_1,\mathbf 1},\cdots, t_{\ee_K,\mathbf 1}}_K,~
 \underbrace{0,\cdots,0}_{J-K}~\big)^\top.
\end{eqnarray*}
By the results that $  t_{\ee_h,\ee_h}  =   t_{\ee_h,\mathbf 1}$ and \eqref{App}, we know  
\begin{eqnarray*}
&& T_{\sum_{k=1}^{K}\ee_k,\Cdot}(Q, \Theta-\ttt^*\mathbf 1^\top)\\
&=&T_{\sum_{k=1}^{K}\ee_k,\Cdot}(Q, \bar\Theta-\ttt^*\mathbf 1^\top) 
~=~\biggr(
\prod_{k=1}^K
( t_{\ee_k,\mathbf 0}- t_{\ee_k,\mathbf 1}),~\mathbf 0^\top\biggr)
\end{eqnarray*}
where the product element is nonzero under the model assumption. Then the equation $$T_{\sum_{k=1}^{K}\ee_k,\Cdot}(Q, \Theta-\ttt^*\mathbf 1^\top)\pp=T_{\sum_{k=1}^{K}\ee_k,\Cdot}(Q, \bar\Theta-\ttt^*\mathbf 1^\top)\bar\pp$$ implies 
$$p_{\mathbf 0} = \bar p_{\mathbf 0}.$$
Now for any $h\in\{1,\cdots, K\}$, take
\begin{eqnarray}\label{theta1}
&&\ttt^* = \big(~\underbrace{ t_{\ee_1,\mathbf 1},\cdots,  t_{\ee_{h-1},\mathbf 1}, t_{\ee_{h},\mathbf 0}, t_{\ee_{h+1},\mathbf 1},\cdots, t_{\ee_K,\mathbf 1}}_K,
~ \underbrace{0,\cdots,0}_{J-K}~\big)^\top.
\end{eqnarray}
From the results in \eqref{App}, we have 
\begin{eqnarray*}
&& T_{\sum_{k=1}^{K}\ee_k,\Cdot}(Q, \Theta-\ttt^*\mathbf 1^\top)
~=~
T_{\sum_{k=1}^{K}\ee_k,\Cdot}(Q, \bar\Theta-\ttt^*\mathbf 1^\top)\\
&&=~
\biggr(\mathbf 0^\top, ~
\underbrace{
( t_{\ee_h,\ee_h}- t_{\ee_h,\mathbf 0})
\prod_{\substack{ k=1,\cdots, K,\\ k\neq h}}
( t_{\ee_k,\ee_h}- t_{\ee_k,\mathbf 1})}_{column~\ee_h},~\mathbf 0^\top\biggr).
\end{eqnarray*}
Then the equation  $T_{\sum_{k=1}^{K}\ee_k,\Cdot}(Q, \Theta-\ttt^*\mathbf 1^\top)\pp=T_{\sum_{k=1}^{K}\ee_k,\Cdot}(Q, \bar\Theta-\ttt^*\mathbf 1^\top)\bar\pp$ implies 
\begin{eqnarray}\label{Apr15_4}
p_{\ee_h} = {\bar p}_{\ee_h} \mbox{ for } h=1,\cdots, K.
\end{eqnarray}

We continue to show $t_{\ee_j,\ee_h} = \bar t_{\ee_j,\ee_h}$
for the rest $j\in\{1,\cdots, 2K\}$ and $h\in\{1,\cdots, K\}$.
Consider any $j$ and $h$ such that $K< j \leq 2K$ and  $1\leq h \leq K$. For 
$\ttt^*$ in \eqref{theta1}
 we have 
\begin{eqnarray*}
&& T_{\ee_j+\sum_{k=1}^{K}\ee_k,\Cdot}(Q, \Theta-\ttt^*\mathbf 1^\top)\\
&=& 
\biggr(\mathbf 0^\top, 
\underbrace{
t_{\ee_j,\ee_h} ( t_{\ee_h,\ee_h}- t_{\ee_h,\mathbf 0})
\prod_{\substack{ k=1,\cdots, K,\\ k\neq h}}
( t_{\ee_k,\ee_h}- t_{\ee_k,\mathbf 1})}_{column~\ee_h},~\mathbf 0^\top\biggr).
\end{eqnarray*}
and 
\begin{eqnarray*}
&& T_{\ee_j+ \sum_{k=1}^{K}\ee_k,\Cdot}(Q, \bar\Theta-\ttt^*\mathbf 1^\top)\\
&=& 
\biggr(\mathbf 0^\top, 
\underbrace{
\bar t_{\ee_j,\ee_h} ( t_{\ee_h,\ee_h}- t_{\ee_h,\mathbf 0})
\prod_{\substack{ k=1,\cdots, K,\\ k\neq h}}
( t_{\ee_k,\ee_h}- t_{\ee_k,\mathbf 1})}_{column~\ee_h},~\mathbf 0^\top\biggr).
\end{eqnarray*}
Then from \eqref{Apr15_4} and  
$$T_{\ee_j+\sum_{k=1}^{K}\ee_k,\Cdot}(Q, \Theta-\ttt^*\mathbf 1^\top)\pp=T_{\ee_j+\sum_{k=1}^{K}\ee_k,\Cdot}(Q, \bar\Theta-\ttt^*\mathbf 1^\top)\bar\pp,$$ we obtain  
$$ t_{\ee_j,\ee_h} = \bar t_{\ee_j,\ee_h}.$$
For any $j$ and $h$ such that $1\leq j \leq K$ and  $1\leq h \leq K$, take
$$
\ttt^* = \big(\underbrace{0,\cdots,0}_{K},
~\underbrace{ t_{\ee_{K+1},\mathbf 1},\cdots,  t_{\ee_{K+h-1},\mathbf 1}, t_{\ee_{K+h},\mathbf 0}, t_{\ee_{K+h+1},\mathbf 1},\cdots, t_{\ee_{2K},\mathbf 1}}_K,
~ \underbrace{0,\cdots,0}_{J-2K}\big)^\top
$$
and a similar argument gives $ t_{\ee_j,\ee_h} = \bar t_{\ee_j,\ee_h}.$
This completes Step 3.



\paragraph{Step 4}
The proof for Step 4 and Step 5 uses similar arguments.
To better illustrate our idea, we separate them in two steps.  In particular, in Step 4,
we consider the columns corresponding to two attributes. 
For any $h_1$ and $h_2$ such that $1\leq h_1< h_2\leq K,$
we first prove $p_{\ee_{h_1}+\ee_{h_2}}=\bar p_{\ee_{h_1}+\ee_{h_2}}.$
Take 
\begin{eqnarray*}
\ttt^* &=&\big (~
\underbrace{ t_{\ee_1,\mathbf 1},\cdots, t_{\ee_{h_1-1},\mathbf 1}, 
 t_{\ee_{h_1},\mathbf 0}}_{h_1}, 
 ~
 \underbrace{ t_{\ee_{h_1+1},\mathbf 1}, \cdots,  t_{\ee_{h_2-1},\mathbf 1}, 
 t_{\ee_{h_2},\ee_{h_1}}}_{h_2-h_1},
\\&&\quad
  \underbrace{t_{\ee_{h_2+1},\mathbf 1}, \cdots, t_{\ee_{K},\mathbf 1}}_
  {K-h_2},~
\underbrace{0,\cdots,0}_{J-K}~\big)^\top.
\end{eqnarray*}
With such a choice of $\ttt^*$,  for any $k\in\{1,\cdots,K\}\backslash\{h_1,h_2\}$,
$ t_{\ee_k,\aaa} - \theta^*_k =  t_{\ee_k,\aaa} -  t_{\ee_k,\mathbf 1}=0$ if $\aaa\succeq \ee_k$.
In addition, $ t_{\ee_{h_2},\ee_{h_1}} - \theta^*_{h_2}=0$. 
Therefore, by the definition, the row vector of $T$-matrix $T(Q, \Theta-\ttt^*\mathbf 1^\top)$ corresponding to $\rr=\sum_{k=1}^{K}\ee_k$ has only two possible nonzero elements, which  correspond to the two columns $\ee_{h_2}$ and $\ee_{h_1}+\ee_{h_2}$ in the $T$-matrix. Specifically, we have  
\begin{align*}
&
\quad T_{\sum_{k=1}^{K}\ee_k,\Cdot}(Q, \Theta-\ttt^*\mathbf 1^\top)\\
&=~\biggr(\mathbf 0^\top,~
\underbrace{ (t_{\ee_{h_1},\ee_{h_2}}-  t_{\ee_{h_1},\mathbf 0})
  (t_{\ee_{h_2},\ee_{h_2}}-  t_{\ee_{h_2},\ee_{h_1}})
  \prod_{\substack{ k=1,\cdots,K, \\
  k\neq h_1,h_2}}( t_{\ee_k,\ee_{h_2}}- t_{\ee_k,\mathbf 1})}
 _{column~\ee_{h_2}},~
\mathbf 0^\top,\\
&
 \underbrace{(t_{\ee_{h_1},\ee_{h_1}+\ee_{h_2}}-  t_{\ee_{h_1},\mathbf 0})
  (t_{\ee_{h_2},\ee_{h_1}+\ee_{h_2}}-  t_{\ee_{h_2},\ee_{h_1}})
  \prod_{\substack{ k=1,\cdots,K, \\
  k\neq h_1,h_2}}
( t_{\ee_k,\ee_{h_1}+\ee_{h_2}}- t_{\ee_k,\mathbf 1})}_{column~\ee_{h_1}+\ee_{h_2}},~\mathbf 0^\top\biggr).
\end{align*}
Consider the row vector of $T$-matrix $T(Q, \bar\Theta-\ttt^*\mathbf 1^\top)$ corresponding to $\rr=\sum_{k=1}^{K}\ee_k$.
Thanks to the results in Steps 1--3, a similar calculation gives the following equation for the chosen $\ttt^*$ 
$$ 
T_{\sum_{k=1}^{K}\ee_k,\Cdot}(Q, \bar\Theta-\ttt^*\mathbf 1^\top)
=T_{\sum_{k=1}^{K}\ee_k,\Cdot}(Q, \Theta-\ttt^*\mathbf 1^\top).
$$
Under the model assumption,  we have 
$$t_{\ee_{h_1},\ee_{h_1}+\ee_{h_2}}-  t_{\ee_{h_1},\mathbf 0}>0,~~
t_{\ee_{h_2},\ee_{h_1}+\ee_{h_2}}-  t_{\ee_{h_2},\ee_{h_1}}>0,
 \prod_{k=1,\cdots,K, \atop k\neq h_1,h_2}( t_{\ee_k,\ee_{h_2}}- t_{\ee_k,\mathbf 1}) \neq 0,$$ and 
$  \prod_{k=1,\cdots,K, k\neq h_1,h_2}
( t_{\ee_k,\ee_{h_1}+\ee_{h_2}}- t_{\ee_k,\mathbf 1})\neq 0$.
Therefore the $\ee_{h_1}+\ee_{h_2}$ column element of  
$T_{\sum_{k=1}^{K}\ee_k,\Cdot}(Q, \bar\Theta-\ttt^*\mathbf 1^\top)$, equivalently $T_{\sum_{k=1}^{K}\ee_k,\Cdot}(Q, \Theta-\ttt^*\mathbf 1^\top)$,
 is nonzero.
From the equation $$T_{\sum_{k=1}^{K}\ee_k,\Cdot}(Q, \Theta-\ttt^*\mathbf 1^\top)\pp
=T_{\sum_{k=1}^{K}\ee_k,\Cdot}(Q, \bar\Theta-\ttt^*\mathbf 1^\top)\bar\pp$$
and the result that $p_{\ee_{h_2}}=\bar p_{\ee_{h_2}}$ as proved in Step 3, 
we thus have 
$$p_{\ee_{h_1}+\ee_{h_2}}=\bar p_{\ee_{h_1}+\ee_{h_2}}.$$

Next we show $  t_{\ee_j ,\ee_{h_1}+\ee_{h_2}} = \bar t_{\ee_j ,\ee_{h_1}+\ee_{h_2}}$.
First consider the case when $j>K$. For  the row vector of $T$-matrix $T(Q, \Theta-\ttt^*\mathbf 1^\top)$ corresponding to $\rr=\sum_{k=1}^{K}\ee_k +\ee_j$, 
we have 
\begin{align}\label{Apr14_6}
&\quad T_{\sum_{k=1}^{K}\ee_k+\ee_j,\Cdot}(Q, \Theta-\ttt^*\mathbf 1^\top)\\
&=~\biggr(\mathbf 0^\top,~
\underbrace{t_{\ee_j ,\ee_{h_2}}
 (t_{\ee_{h_1},\ee_{h_2}}-  t_{\ee_{h_1},\mathbf 0})
  (t_{\ee_{h_2},\ee_{h_2}}-  t_{\ee_{h_2},\ee_{h_1}})
  \prod_{\substack{ k=1,\cdots,K, \\
  k\neq h_1,h_2}}( t_{\ee_k,\ee_{h_2}}- t_{\ee_k,\mathbf 1})}
_{column~\ee_{h_2}},
~\mathbf 0^\top,~\notag\\
&
 \underbrace{t_{\ee_j ,\ee_{h_1}+\ee_{h_2}}
 (t_{\ee_{h_1},\ee_{h_1}+\ee_{h_2}}-  t_{\ee_{h_1},\mathbf 0})
  (t_{\ee_{h_2},\ee_{h_1}+\ee_{h_2}}-  t_{\ee_{h_2},\ee_{h_1}})
  \prod_{\substack{ k=1,\cdots,K, \notag\\
  k\neq h_1,h_2}}
( t_{\ee_k,\ee_{h_1}+\ee_{h_2}}- t_{\ee_k,\mathbf 1})
 }_{column~\ee_{h_1}+\ee_{h_2}},\notag\\
 &~\mathbf 0^\top\biggr).\notag
\end{align}
Similarly, for  the row vector of $T$-matrix $T(Q, \bar\Theta-\ttt^*\mathbf 1^\top)$ corresponding to $\rr=\sum_{k=1}^{K}\ee_k +\ee_j$,  we can write 
\begin{align}\label{Apr14_7}
&\quad T_{\sum_{k=1}^{K}\ee_k+\ee_j,\Cdot}(Q, \bar\Theta-\ttt^*\mathbf 1^\top)
\\
&=\,\biggr(\mathbf 0^\top,
\underbrace{ t_{\ee_j ,\ee_{h_2}}
 (t_{\ee_{h_1},\ee_{h_2}}-  t_{\ee_{h_1},\mathbf 0})
  (t_{\ee_{h_2},\ee_{h_2}}-  t_{\ee_{h_2},\ee_{h_1}})
   \prod_{\substack{ k=1,\cdots,K, \notag\\
  k\neq h_1,h_2}}( t_{\ee_k,\ee_{h_2}}- t_{\ee_k,\mathbf 1})}_{column~\ee_{h_2}},
~\mathbf 0^\top,\\
&
 \underbrace{\bar t_{\ee_j ,\ee_{h_1}+\ee_{h_2}}
 (t_{\ee_{h_1},\ee_{h_1}+\ee_{h_2}}-  t_{\ee_{h_1},\mathbf 0})
  (t_{\ee_{h_2},\ee_{h_1}+\ee_{h_2}}-  t_{\ee_{h_2},\ee_{h_1}})
  \prod_{\substack{ k=1,\cdots,K,\notag \\
  k\neq h_1,h_2}}
( t_{\ee_k,\ee_{h_2}}- t_{\ee_k,\mathbf 1})
 }_{column~\ee_{h_1}+\ee_{h_2}},\notag\\
 &~\mathbf 0^\top\biggr),\notag
\end{align}
where  the result  $ \bar t_{\ee_j ,\ee_{h_2}} = t_{\ee_j ,\ee_{h_2}}$ is used for the element in  column $\ee_{h_2}$.
From \eqref{Apr14_6}, \eqref{Apr14_7}, and the proved results that $p_{\ee_{h_2}} =\bar p_{\ee_{h_2}}$ and $p_{\ee_{h_1}+\ee_{h_2}}=\bar p_{\ee_{h_1}+\ee_{h_2}}$, we can derive   
$$  t_{\ee_j ,\ee_{h_1}+\ee_{h_2}} = \bar t_{\ee_j ,\ee_{h_1}+\ee_{h_2}},$$
for any $1\leq h_1< h_2\leq K$ and $j>K$,
from the equation $T_{\sum_{k=1}^{K}\ee_k+\ee_j,\Cdot}(Q, \Theta-\ttt^*\mathbf 1^\top)\pp
=T_{\sum_{k=1}^{K}\ee_k+\ee_j,\Cdot}(Q, \bar\Theta-\ttt^*\mathbf 1^\top)\bar\pp.$

Moreover, for any $1\leq J\leq K$ and $1\leq h_1< h_2\leq K$, we redefine  
\begin{align*}
\ttt^* &=~ \big(\underbrace{0,\cdots,0}_{K}, 
~\underbrace{t_{\ee_{K+1},\mathbf 1},\cdots, t_{\ee_{K+h_1-1},\mathbf 1}, 
 t_{\ee_{K+h_1},\mathbf 0}}_{h_1}, 
\\&
\underbrace{ t_{\ee_{K+h_1+1},\mathbf 1}, \cdots,  t_{\ee_{K+h_2-1},\mathbf 1}, 
 t_{\ee_{K+h_2},\ee_{h_1}}}_{h_2-h_1},~
 \underbrace{  t_{\ee_{K+h_2+1},\mathbf 1}, \cdots,  t_{\ee_{2K},\mathbf 1}}_{K-h_2},
~\underbrace{0,\cdots,0}_{J-2K}~\big)^\top.
\end{align*}
Consider $T_{\sum_{k=K+1}^{2K}\ee_k,\Cdot}(Q, \Theta-\ttt^*\mathbf 1^\top)$ instead of $T_{\sum_{k=1}^{K}\ee_k,\Cdot}(Q, \Theta-\ttt^*\mathbf 1^\top)$. A similar argument as above gives  
$$  t_{\ee_j ,\ee_{h_1}+\ee_{h_2}} = \bar t_{\ee_j ,\ee_{h_1}+\ee_{h_2}}$$
for any $1\leq h_1< h_2\leq K$ and $j=1,\cdots, K$.
This completes Step 4.

\paragraph{Step 5} 
We consider the columns corresponding to more than two attributes. 
We use the induction method and a similar argument as in Step 4.
In particular, consider any integer $k$ such that $3\leq k\leq K$. For any $l\leq k-1$,
suppose we have 
$$t_{\ee_j,\sum_{i=1}^{l}\ee_{h_i}}=\bar t_{\ee_j,\sum_{i=1}^{l}\ee_{h_i}}
\mbox{ and } p_{\sum_{i=1}^{l}\ee_{h_i}}=\bar p_{\sum_{i=1}^{l}\ee_{h_i}}$$
for any $j\in\{1,\cdots,J\}$ and  $1\leq h_1,\cdots,h_l\leq K$. 
We next show that the two equations also hold for $l=k$. 

Consider any  $1\leq h_1,\cdots,h_k\leq K$.
  Define the vector  $\ttt^*=(\theta^*_1,\cdots, \theta^*_J)^\top$ as
$$
\theta^*_{i}  =
 \begin{cases}
 t_{\ee_{i}, \mathbf 0} & \mbox{for } i \in \{h_1,\cdots,h_k\}; \\
 t_{\ee_{i}, \mathbf 1} &  \mbox{for } i\in \{1,\cdots,K\}\setminus\{h_1,\cdots,h_k\};\\
0 & \mbox{otherwise}.
\end{cases}
$$
Then under the induction assumption, we have the equivalence of the two row vectors:
$$T_{\sum_{i=1}^{K}\ee_{i},\Cdot}(Q,\Theta-\ttt^*\mathbf 1^\top)
=T_{\sum_{i=1}^{K}\ee_{i},\Cdot}(Q,\bar\Theta-\ttt^*\mathbf 1^\top).$$
In particular,
the element of $T_{\sum_{i=1}^{K}\ee_{i},\Cdot}(Q,\Theta-\ttt^*\mathbf 1^\top)$ 
corresponding to column $\sum_{i=1}^{k}\ee_{h_i}$ is nonzero; 
for any $l<k$, the elements corresponding to column $\sum_{i=1}^{l}\ee_{h_i}$ may be zero or nonzero;
and the others terms are 0.
Since $p_{\sum_{i=1}^{l}\ee_{h_i}}=\bar p_{\sum_{i=1}^{l}\ee_{h_i}}$ for any $l<k$,  the equation $T_{\sum_{i=1}^{k}\ee_{h_i},\Cdot}(Q,\Theta)\pp =T_{\sum_{i=1}^{k}\ee_{h_i},\Cdot}(Q,\bar\Theta)\bar\pp $
gives 
$$p_{\sum_{i=1}^{k}\ee_{h_i}}=\bar p_{\sum_{i=1}^{k}\ee_{h_i}}.$$
Moreover,  for any $j>K$, we have $T_{\ee_j+\sum_{i=1}^{K}\ee_{i},\Cdot}(Q,\Theta)\pp = T_{\ee_j+\sum_{i=1}^{K}\ee_{i},\Cdot}(Q,\bar\Theta)\bar\pp $. Following  a similar argument as in Step 4, we can establish  
$$t_{\ee_j,\sum_{i=1}^{k}\ee_{h_i}}=\bar t_{\ee_j,\sum_{i=1}^{k}\ee_{h_i}}.$$
For $1\leq j\leq K$ and $1\leq h_1,\cdots,h_k\leq K$, take 
$$
\theta^*_{i}  =
 \begin{cases}
 t_{\ee_{i}, \mathbf 0} & \mbox{for } i \in \{K+h_1,\cdots,K+h_k\}; \\
 t_{\ee_{i}, \mathbf 1} &  \mbox{for } i\in \{K+1,\cdots,2K\}\setminus\{K+h_1,\cdots,K+h_k\};\\
0 & \mbox{otherwise}.
\end{cases}
$$ Similarly we can obtain $t_{\ee_j,\sum_{i=1}^{k}\ee_{h_i}}=\bar t_{\ee_j,\sum_{i=1}^{k}\ee_{h_i}}.$
This completes the proof. 

\subsection{Proofs of Propositions 2--3 and Lemmas 1--2}

\begin{proof}[Proof of the Proposition \ref{theorem0}]
We only need to show that there exist  $(\Theta,\pp)\neq (\bar\Theta,\bar\pp)$ satisfying equation \eqref{mleeq}.
For notational convenience, we write $t_{\ee_j,\aaa}(Q,\Theta)$ and  
$t_{\ee_j,\aaa}(Q,\bar\Theta)$
 as $t_{\ee_j,\aaa}$ and $\bar t_{\ee_j,\aaa}$, respectively. 

For simplicity, consider the DINA model in Example 2, under which 
$t_{\ee_j,\aaa} = t_{\ee_j,\mathbf 0}$ if $\xi_{j,\aaa}=0$ and 
$t_{\ee_j,\aaa} = t_{\ee_j,\mathbf 1}$ if $\xi_{j,\aaa}=1$.
Without loss of generality, we focus on the $Q$-matrix has
the following form:
$$
Q=
\left(\begin{array}{cc}
1 & \mathbf{0}^{\top}  \\
1 & \mathbf{0}^{\top}  \\
0 & {\cal I}_{K-1}\\
0 & {\cal I}_{K-1}\\
\mathbf{0} & Q^*
\end{array}\right),
$$
where $Q^*$ is unspecified. Note that the above $Q$-matrix does not satisfy condition \ref{C2} under the DINA model.
Next we show the item parameters for the first two items are non-identifiable. 

Let  $t_{\ee_j,\mathbf 1} = \bar t_{\ee_j,\mathbf 1}$ for $j\geq 3$.
Consider the row vector of the $T$-matrix corresponding to
 $\rr=(r_1, r_2, \cdots, r_J)^\top$.
Consider each possible value of $(r_1,r_2) \in \{0,1\}^2$. We can show that for any  $(\Theta,\pp)\neq (\bar\Theta,\bar\pp)$, 
equation \eqref{mleeq} is satisfied if 
the following equations hold for any $\aalpha\in\{0,1\}^K$ such that $\alpha_1 = 0$:
\begin{equation}\label{eq:Proof3.1}
\begin{cases}
 p_\aalpha +  p_{\aalpha+\ee_1} 
=\bar p_\aalpha + \bar p_{\aalpha+\ee_1}, 
& \mbox{ if } (r_1,r_2) = (0,0);\\ 
t_{\ee_1,\mathbf 1}  p_{\aalpha+\ee_1}+ t_{\ee_1,\mathbf 0}  p_{\aalpha} 
= 
\bar t_{\ee_1,\mathbf 1} \bar p_{\aalpha+\ee_1}+\bar t_{\ee_1,\mathbf 0} \bar p_{\aalpha},
 &  \mbox{ if } (r_1,r_2) = (1,0);\\ 
 t_{\ee_2,\mathbf 1}  p_{\aalpha+\ee_1}+  t_{\ee_2,\mathbf 0}  p_{\aalpha}
=
\bar t_{\ee_2,\mathbf 1} \bar p_{\aalpha+\ee_1}
+\bar t_{\ee_2,\mathbf 0}  \bar p_{\aalpha} ,
 & \mbox{ if }  (r_1,r_2) = (0,1);\\ 
t_{\ee_1,\mathbf 1}  t_{\ee_2,\mathbf 1}  
 p_{\aalpha+\ee_1}
+ t_{\ee_1,\mathbf 0}  t_{\ee_2,\mathbf 0}   p_{\aalpha}
\\
\quad\quad \quad\quad
=\bar t_{\ee_1,\mathbf 1} \bar t_{\ee_2,\mathbf 1}   \bar p_{\aalpha+\ee_1}
+\bar  t_{\ee_1,\mathbf 0} \bar t_{\ee_2,\mathbf 0} \bar p_{\aalpha},
& \mbox{ if }  (r_1,r_2) = (1,1).
\end{cases}
\end{equation}

Now we construct $(\Theta,\pp)\neq(\bar\Theta,\bar\pp)$ such that \eqref{eq:Proof3.1} is satisfied. For $\rho\in(0,1)$,
choose $(\Theta,\pp)$ such that $ p_\aalpha /  p_{\aalpha+\ee_1} = \rho$ for over all $\aalpha \in \{0,1\}^K$ with $\alpha_1 =0$. 
Then, for any $\bar t_{\ee_j,\mathbf 0}$, $j=1,\cdots, J$, define 
\begin{align*}
\bar t_{\ee_j,\mathbf 1} =&~\left\{ \begin{array}{r}
\bar t_{\ee_1,\mathbf 0} + \resizebox{7.5cm}{!}{$\frac{( t_{\ee_1,\mathbf 1}- \bar t_{\ee_1,\mathbf 0}) ( t_{\ee_2,\mathbf 1}-\bar  t_{\ee_2,\mathbf 0}) + \rho ( t_{\ee_1,\mathbf 0}-\bar  t_{\ee_1,\mathbf 0}) (t_{\ee_2,\mathbf 0}-\bar  t_{\ee_2,\mathbf 0})}
{(  t_{\ee_2,\mathbf 1}-\bar  t_{\ee_2,\mathbf 0}) + \rho ( t_{\ee_2,\mathbf 0} -\bar t_{\ee_2,\mathbf 0} )}$}, \mbox{ if } j = 1;
\vspace{0.1in}\\
\bar t_{\ee_2,\mathbf 0}  + 
\resizebox{7.5cm}{!}{$\frac{( t_{\ee_1,\mathbf 1} -\bar t_{\ee_1,\mathbf 0} ) ( t_{\ee_2,\mathbf 1} -\bar t_{\ee_2,\mathbf 0} ) + \rho ( t_{\ee_1,\mathbf 0} -\bar t_{\ee_1,\mathbf 0} ) ( t_{\ee_2,\mathbf 0} -\bar t_{\ee_2,\mathbf 0} )}{( t_{\ee_1,\mathbf 1} -\bar t_{\ee_1,\mathbf 0} ) + \rho ( t_{\ee_1,\mathbf 0} -\bar t_{\ee_1,\mathbf 0} )}$},  \mbox{ if } j = 2; 
\\
 t_{\ee_j,\mathbf 1},\quad\quad \quad\quad \quad\quad \quad\quad 
 \quad\quad \quad\quad \quad\quad \quad\quad \quad\quad\: 
  \mbox{if }  j =3, \ldots, J;
\end{array}
\right.\\
\bar p_{\aalpha+\ee_1} =&~
\resizebox{8cm}{!}{$  \frac{\left\{( t_{\ee_1,\mathbf 1} -\bar t_{\ee_1,\mathbf 0} ) + \rho ( t_{\ee_1,\mathbf 0} -\bar t_{\ee_1,\mathbf 0} )\right\}
  \left\{( t_{\ee_2,\mathbf 1} -\bar t_{\ee_2,\mathbf 0} ) + \rho ( t_{\ee_2,\mathbf 0}  - \bar t_{\ee_2,\mathbf 0} ) \right\}}
  {( t_{\ee_1,\mathbf 1} -\bar t_{\ee_1,\mathbf 0} )(  t_{\ee_2,\mathbf 1} -\bar t_{\ee_2,\mathbf 0} ) + \rho ( t_{\ee_1,\mathbf 0} -\bar t_{\ee_1,\mathbf 0} )(  t_{\ee_2,\mathbf 0} -\bar t_{\ee_2,\mathbf 0} )}$} \times
  p_{\aalpha+\ee_1},\\
\bar p_\aalpha =&~  p_\aalpha +  p_{\aalpha+\ee_1} - \bar p_{\aalpha+\ee_1},
\end{align*}
for every $\aalpha \in \{0,1\}^K$ such that $\alpha_1=0$.
This results in a solution to \eqref{eq:Proof3.1}. 
Thus, we have constructed $(\Theta,\pp)\neq(\bar\Theta,\bar\pp)$ such that  \eqref{mleeq} holds. This completes the proof.
\end{proof}

\begin{proof}[Proof of the Proposition~\ref{trans}]
In what follows, we construct  a $D$ matrix satisfying the conditions in the proposition, i.e.,  $D(\ttt^*)$ is a matrix
only depending on $\ttt^*$ such that $D(\ttt^*) T(Q,\Theta) =T(Q, \Theta-\ttt^*\mathbf 1^\top)$ for any $Q$ and $\Theta$. 
Recall that 
$$t_{\rr,\aalpha}(Q,\Theta)
=\prod_{j: r_j =1} t_{\ee_j,\aalpha}(Q,\Theta),
\ \forall\ \rr \in \{0,1\}^J, \aalpha\in\{0,1\}^K.$$
For any $\ttt^*=(\theta^*_1,\cdots, \theta^*_J) \in \mathbb{R}^J$, 
\begin{eqnarray*}
t_{\rr,\aalpha}(Q,\Theta-\ttt^*\mathbf 1^\top)
= \prod_{j:r_j=1} \{t_{\ee_j,\aalpha}(Q,\Theta) -\theta_j^*\}.
\end{eqnarray*}
By polynomial expansion,
\begin{eqnarray*}
t_{\rr,\aalpha}(Q,\Theta-\ttt^*\mathbf 1^\top)
= \sum_{\rr' \preceq \rr} (-1)^{\sum_{j=1}^J r_j-r_j'}\prod_{j: r_j - r_j' = 1} \theta_j^* 
\prod_{k: r_k' = 1} t_{\ee_k,\aalpha}(Q,\Theta).
\end{eqnarray*}
Define the entrie $d_{\rr,\rr'}(\ttt^*)$ of $D(\ttt^*)$ corresponding to row $\rr$ and column 
$\rr'$ as
$$
d_{\rr,\rr'}(\ttt^*) = \begin{cases}
0 & \rr' \not\preceq \rr \\
(-1)^{\sum_{j=1}^J r_j-r_j'}\prod_{j: r_j - r_j' = 1} \theta_j^* & \rr' \preceq \rr \mbox{ and } \rr'\neq \rr \\
1 & \rr' = \rr
\end{cases}.
$$
Then we have
$$T(Q,\Theta-\ttt^*\mathbf 1^\top) = D(\ttt^*)T(Q,\Theta),$$
where $D(\ttt^*)$ is a lower triangular matrix  depending solely on $\ttt^*$ with eigenvalues equal to its diagonal. Since $\text{diag}\{D(\ttt^*)\}  = \bfone$, $D(\ttt^*)$ is invertible.
\end{proof}

\begin{proof}[Proof of Lemma \ref{lemma0}]
We use the method of contradiction.
If there exists $k \in\{1,\cdots,K\}$ such that $t_{\ee_k,\mathbf 0}=\bar t_{\ee_k,\aaa^*}$ with $\aaa^*\succeq\ee_k$.
 Since $t_{\ee_k,\mathbf 0}\leq t_{\ee_k,\aaa}$ for any $\aaa\in\{0,1\}^K$ and $t_{\ee_k,\mathbf 0}< t_{\ee_k,\aaa^*} = t_{\ee_k,\mathbf 1}$, this implies that for the row vectors corresponding to 
$\rr =\ee_k$, 
$$T_{\ee_k,\Cdot}(Q, \Theta)\pp
> \sum_\aaa \pp_\aaa t_{\ee_k,\mathbf 0}
=  \sum_\aaa \bar\pp_\aaa \bar t_{\ee_k,\aaa^*}
=  \sum_\aaa \bar\pp_\aaa \bar t_{\ee_k,\mathbf 1}
> T_{\ee_k,\Cdot}(Q, \bar\Theta)\bar\pp,
$$
which contradicts the equation \eqref{mleeq} that requires 
$T_{\ee_k,\Cdot}(Q, \Theta)\pp=T_{\ee_k,\Cdot}(Q, \bar\Theta)\bar\pp.$
Therefore we conclude that $t_{\ee_k,\mathbf 0}\neq\bar t_{\ee_k,\aaa^*}$.
Similarly, we have $ t_{\ee_k,\aaa^*}\neq\bar t_{\ee_k,\mathbf 0}, t_{\ee_{K+k},\mathbf 0}\neq \bar t_{\ee_{K+k},\aaa^*} 
\mbox{ and }
t_{\ee_{K+k},\aaa^*}\neq \bar t_{\ee_{K+k},\mathbf 0}.$
\end{proof}

\begin{proof}[Proof of Lemma \ref{lemma1}]
Without loss of generality, we only need to show that for any $1\leq h\leq K$,
$t_{\ee_1, \ee_h}\neq \bar t_{\ee_1, \mathbf 1}$.

Take
$$\ttt^*=\big(~
\underbrace{ t_{\ee_{1},\mathbf 0},  t_{\ee_{2},\mathbf 1},
\cdots, t_{\ee_{K},\mathbf 1}}_K, ~
\underbrace{0,\cdots,0}_{J-K}~\big)^\top,$$
and we have
\begin{eqnarray*}
T_{\sum_{k=1}^{K}\ee_k,\Cdot}(Q, \Theta-\ttt^*\mathbf 1^\top)
=\biggr(0,~
( t_{\ee_{1},\ee_{1}}- t_{\ee_{1},\mathbf 0})\times
\prod_{k=2}^{K}( t_{\ee_k,\ee_{1}}- t_{\ee_k,\mathbf 1}),
~\mathbf 0^\top\biggr).
\end{eqnarray*}
From the model assumption, the product element is nonzero. 

Consider the row vector $T_{\sum_{k=1}^{K}\ee_k,\Cdot}(Q, \bar\Theta-\ttt^*\mathbf 1^\top)$. Under the equation \eqref{mleeq}, there must exist a nonzero element. We denote the corresponding column as $\aaa^*$ and the element then can be written as 
$$( \bar t_{\ee_{1},\aaa^*}- t_{\ee_{1},\mathbf 0})\times
\prod_{k=2}^{K}(\bar t_{\ee_k,\aaa^*}- t_{\ee_k,\mathbf 1})
\neq 0. $$
Note that here we do not know whether $\aaa^*$ equals $\ee_1$.

Denote $Q_1$ as the $Q$-matrix corresponding to items from $K+1$ to $2K$. 
Note that $Q_1={\cal I}_K$.
Consider the $2^K\times 2^K$ $T$-matrix, $T(Q_1,\bar\Theta_{(K+1):2K})$, where 
$\bar\Theta_{(K+1):2K}$ denotes the submatrix of $\Theta$ containing rows from $K+1$ to $2K$. Take $\tilde \ttt =(\bar \theta_{K+1,\mathbf 1},\cdots,\bar\theta_{2K,\mathbf 1})^\top$, and we know the transformed $T$-matrix $T(Q_1,\bar\Theta_{(K+1):2K}-\tilde \ttt\mathbf 1^\top)$ takes an upper-left triangular form (up to column swapping) and therefore is full rank.
This implies $T(Q_1,\bar\Theta_{(K+1):2K})$ is full rank and thus there exists a row vector $\mm$ such that 
$$\mm \cdot T(Q_1,\bar\Theta_{(K+1):2K}) = (0,\cdots,0,~ \underbrace{~1~}_{column~ \aaa^*},~0,\cdots,0).$$
On the other hand, consider  $\mm\cdot T(Q_1,\Theta_{K+1:2K})$.
We use $x$ to denote the element  corresponding to the column $\ee_1$ (i.e., the second element).
Combining the above results, we know 
\begin{eqnarray*}
&&\{\mm \cdot T(Q_1,\Theta)\}\odot T_{\sum_{k=1}^{K}\ee_k,\Cdot}(Q, \Theta-\ttt^*\mathbf 1^\top) \\
&=& \left(0, x\times( t_{\ee_{1},\ee_{1}}- t_{\ee_{1},\mathbf 0})\times
\prod_{k=2}^{K}( t_{\ee_k,\ee_{1}}- t_{\ee_k,\mathbf 1}), \mathbf 0\right);\\
\mbox{and}&&\{\mm \cdot T(Q_1,\bar\Theta)\}\odot T_{\sum_{k=1}^{K}\ee_k,\Cdot}(Q, \bar\Theta-\ttt^*\mathbf 1^\top)\\
&=& \biggr(0,\cdots,0,~ \underbrace{( \bar t_{\ee_{1},\aaa^*}- t_{\ee_{1},\mathbf 0})\times
\prod_{k=2}^{K}(\bar t_{\ee_k,\aaa^*}- t_{\ee_k,\mathbf 1})}_{column~ \aaa^*},~0,\cdots,0\biggr).
\end{eqnarray*}
Under the equation \eqref{mleeq}, we know $x\neq 0$ and the above two vectors are both nonzero.
Now consider $j>2K$, and we have
\begin{eqnarray*}
&&\{\mm \cdot T(Q_1,\Theta)\}\odot T_{\ee_j+\sum_{k=1}^{K}\ee_k,\Cdot}(Q, \Theta-\ttt^*\mathbf 1^\top) \\
&=& \left(0, x\times  t_{\ee_{j},\ee_{1}}\times ( t_{\ee_{1},\ee_{1}}- t_{\ee_{1},\mathbf 0})\times
\prod_{k=2}^{K}( t_{\ee_k,\ee_{1}}- t_{\ee_k,\mathbf 1}), \mathbf 0\right);\\
\mbox{and}&&\{\mm \cdot T(Q_1,\bar\Theta)\}\odot T_{\ee_j+\sum_{k=1}^{K}\ee_k,\Cdot}(Q, \bar\Theta-\ttt^*\mathbf 1^\top)\\
&=& \biggr(0,\cdots,0,~ \underbrace{
\bar t_{\ee_{j},\aaa^*}\times 
( \bar t_{\ee_{1},\aaa^*}- t_{\ee_{1},\mathbf 0})\times
\prod_{k=2}^{K}(\bar t_{\ee_k,\aaa^*}- t_{\ee_k,\mathbf 1})}_{column~ \aaa^*},~0,\cdots,0\biggr).
\end{eqnarray*}
Therefore as in Step 1, we have for $j>2K$,
$t_{\ee_j,\ee_1} 
= \bar t_{\ee_j,\aaa^*}.$

Now redefine $\ttt^*=(~
\underbrace{0,  t_{\ee_{2},\mathbf 1},
\cdots, t_{\ee_{K},\mathbf 1}}_K, ~
\underbrace{0,\cdots,0}_{J-K}~)^\top,$
and we have
\begin{eqnarray*}
&&T_{\sum_{k=2}^{K}\ee_k,\Cdot}(Q, \Theta-\ttt^*\mathbf 1^\top)\\
&=&\biggr(
\prod_{k=2}^{K}( t_{\ee_k,\bfzero}- t_{\ee_k,\mathbf 1}),
~
\prod_{k=2}^{K}( t_{\ee_k,\ee_{1}}- t_{\ee_k,\mathbf 1}),
~\mathbf 0^\top\biggr),\\
&&T_{\sum_{k=1}^{K}\ee_k,\Cdot}(Q, \Theta-\ttt^*\mathbf 1^\top)\\
&=&\biggr(t_{\ee_{1},\bfzero}
\prod_{k=2}^{K}( t_{\ee_k,\bfzero}- t_{\ee_k,\mathbf 1}),
~
t_{\ee_{1},\ee_{1}}
\prod_{k=2}^{K}( t_{\ee_k,\ee_{1}}- t_{\ee_k,\mathbf 1}),
~\mathbf 0^\top\biggr).
\end{eqnarray*}
From the model assumption, the product elements are nonzero. 
Following the notation in Step 3, there exists a $(J-2K+1)$-dimensional  vector $\uu_1$ such that 
$$b_1=\uu_1 (1, t_{\ee_{2K+1},\ee_1}, \cdots, t_{\ee_{J},\ee_1})^\top\neq0 \mbox{ and }
\uu_1 ( 1, t_{\ee_{2K+1},\mathbf 0}, \cdots, t_{\ee_{J},\mathbf 0})^\top=0.$$
Since  for $j>2K$,
$t_{\ee_j,\ee_1} 
= \bar t_{\ee_j,\aaa^*},$ from a similar argument in Step 3, we have
\begin{eqnarray*}
&&(\uu_1 A)\odot\{\mm \cdot T(Q_1,\Theta)\}\odot T_{\sum_{k=2}^{K}\ee_k,\Cdot}(Q, \Theta-\ttt^*\mathbf 1^\top) \\
&=& \biggr(0, b_1\times x\times
\prod_{k=2}^{K}( t_{\ee_k,\ee_{1}}- t_{\ee_k,\mathbf 1}), \mathbf 0^\top\biggr);\\
&&(\uu_1 A)\odot\{\mm \cdot T(Q_1,\Theta)\}\odot T_{\sum_{k=1}^{K}\ee_k,\Cdot}(Q, \Theta-\ttt^*\mathbf 1^\top) \\
&=& \biggr(0, b_1\times x\times t_{\ee_{1},\ee_{1}}\times
\prod_{k=2}^{K}( t_{\ee_k,\ee_{1}}- t_{\ee_k,\mathbf 1}), \mathbf 0^\top\biggr);\\
&&
(\uu_1 \bar  A)\odot\{\mm \cdot T(Q_1,\bar\Theta)\}\odot T_{\sum_{k=2}^{K}\ee_k,\Cdot}(Q, \bar\Theta-\ttt^*\mathbf 1^\top)\\
&=& \biggr(0,\cdots,0,~ \underbrace{b_1\times
\prod_{k=2}^{K}(\bar t_{\ee_k,\aaa^*}- t_{\ee_k,\mathbf 1})}_{column~ \aaa^*},~0,\cdots,0\biggr);\\
\mbox{and}&&
(\uu_1 \bar A)\odot\{\mm \cdot T(Q_1,\bar\Theta)\}\odot T_{\sum_{k=1}^{K}\ee_k,\Cdot}(Q, \bar\Theta-\ttt^*\mathbf 1^\top)\\
&=& \biggr(0,\cdots,0,~ \underbrace{b_1\times \bar t_{\ee_{1},\aaa^*}\times
\prod_{k=2}^{K}(\bar t_{\ee_k,\aaa^*}- t_{\ee_k,\mathbf 1})}_{column~ \aaa^*},~0,\cdots,0\biggr).
\end{eqnarray*}
The above equations imply that $t_{\ee_{1},\ee_{1}}= \bar t_{\ee_{1},\aaa^*}.$
Since under the model assumption 
$t_{\ee_{1},\ee_{1}}> t_{\ee_1, \ee_h}$, we have the conclusion that 
$ t_{\ee_1, \ee_h}\neq \bar t_{\ee_{1},\mathbf 1}$ since otherwise, we have 
$ \bar t_{\ee_{1},\aaa^*}>\bar t_{\ee_{1},\mathbf 1}$ which cannot be true under the model assumption.
This completes the proof.
\end{proof}

\section*{Acknowledgment}
 The author thanks the editor, the associate editor, and three reviewers for many helpful and constructive comments.

\bibliographystyle{asa}
\bibliography{bibEduc}

\end{document}